\documentclass[a4paper,onecolumn,allowtoday,unpublished]{quantumarticle}
\pdfoutput=1
\usepackage[T1]{fontenc}
\usepackage[british]{babel}
\usepackage[unicode,colorlinks]{hyperref}
\usepackage{graphicx}
\usepackage{subcaption}
\usepackage{amsmath,amssymb,amsthm,bm}
\usepackage{xspace}
\usepackage{mathtools}
\usepackage{dsfont}
\usepackage{xcolor}
\usepackage{tikz,pgfplots}\pgfplotsset{compat=1.18}

\pgfplotsset{
    legend image with text/.style={
        legend image code/.code={%
            \node[anchor=center] at (0.3cm,0cm) {#1};
        }
    },
}

\usepackage{booktabs}
\usepackage{tabularx}

\usepackage{mathrsfs}
\usepackage[all]{xy}

\usepackage{csquotes}
\usepackage[backend=bibtex,style=phys,biblabel=brackets,eprint=true,doi=false,url=true,maxnames=10,giveninits=true]{biblatex}
\usepackage{changes}
\DeclareMathOperator{\cl}{cl}

\let\originalleft\left
\let\originalright\right
\renewcommand{\left}{\mathopen{}\mathclose\bgroup\originalleft}
\renewcommand{\right}{\aftergroup\egroup\originalright}

\newcommand{\bra}[1]{\left\langle #1 \right|}
\newcommand{\ket}[1]{\left| #1 \right\rangle}
\newcommand{\braket}[2]{\left\langle #1 \middle| #2 \right\rangle}
\newcommand{\ketbra}[2]{\left|#1\middle\rangle\middle\langle#2\right|}
\newcommand{\proj}[1]{\left|#1\middle\rangle\middle\langle#1\right|}

\newcommand{\norm}[1]{\left\|#1\right\|}

\newcommand{\id}{\mathds{1}}

\newcommand{\R}{\mathbb{R}}

\newcommand{\mmeas}[0]{\widetilde{\mathcal{M}}}

\newtheorem{theorem}{Theorem}[section]
\newtheorem*{theorem*}{Theorem}
\newtheorem{lemma}[theorem]{Lemma}
\newtheorem{proposition}[theorem]{Proposition}

\def\Pr{\operatorname{Pr}}
\def\Tr{\operatorname{Tr}}

\newcommand{\g}{\mathcal G}
\newcommand{\z}{\mathcal Z}
\newcommand{\zg}{\mathcal Z \circ \mathcal G}

\newcommand{\gmap}{\widehat{\mathcal G}}

\newcommand{\zgmap}{\widehat{\mathcal{ZG}}}
\newcommand{\pkey}{p^\mathrm{K}}

\newcommand{\Psihat}{\widehat\Psi_\mu}

\newcommand{\pro}[1]{\ket{#1}\bra{#1}}
\newcommand{\mc}[1]{\mathcal{#1}}

\newtheorem{remark}{Remark}

\begin{document}

\title{
Effective discrete-modulated continuous variable QKD under general attacks using dimension reduction}

\author{Mariana Navarro }
\orcid{0000-0001-9381-369X}
\email{mariana.navarro@icfo.eu}
\affiliation{Luxquanta Technologies S.L., Av. Joan Carles I, 30, 1º1ª. 08908 L´Hospitalet de Llobregat, Barcelona, Spain}
\affiliation{ICFO - Institut de Ciencies Fotoniques, The Barcelona Institute of Science and Technology, \\ 08860 Castelldefels, Barcelona, Spain}

\author{Mateus Araújo}
\orcid{0000-0003-0155-354X}
\affiliation{Departamento de Física Teórica, Atómica y Óptica, Laboratory for Disruptive Interdisciplinary Science (LaDIS), Universidad de Valladolid, 47011 Valladolid, Spain}

\author{Antonio Acín}
\orcid{0000-0002-1355-3435}
\affiliation{ICFO - Institut de Ciencies Fotoniques, The Barcelona Institute of Science and Technology, \\ 08860 Castelldefels, Barcelona, Spain}
\affiliation{ICREA – Institució Catalana de Recerca i Estudis Avançats, Lluís Companys 23, 08010 Barcelona, Spain}

\author{Carlos Pascual-García } 
\orcid{0000-0003-0659-7349}
\affiliation{Luxquanta Technologies S.L., Av. Joan Carles I, 30, 1º1ª. 08908 L´Hospitalet de Llobregat, Barcelona, Spain}

\begin{abstract}
    Continuous variable quantum key distribution via discrete modulations ensures information-theoretic security using standard telecom technologies, providing affordable and scalable
    quantum communications with simplified classical postprocessing.
    However,  existing security proofs against general attacks often rely on restrictive assumptions, such as a bounded dimension for coherent states, or require impractically large block sizes. In this work, we develop a finite-size security analysis that removes these limitations while incorporating realistic experimental features. Our approach combines an optimization under infinite dimensions, a security proof against general attacks, a trusted detector model accounting for the receiver imperfections, and postselection strategies to reuse discarded measurement outcomes. 
    We report positive variable-length key rates in the finite-size regime for relevant block sizes of the order of $10^6$. These results contribute to narrowing the gap between theoretical security proofs and practical implementations of discrete-modulated continuous variable quantum key distribution protocols.
    \end{abstract}

\maketitle

% FALTA POR REVISAR
% - Intro
% - Conclu
% - Seccion resultados
% - Apéndice H

\section{Introduction}
% Usual introduction of QKD
Quantum key distribution (QKD) \cite{BB84,BBM92,E91} enables private and secure communication between two honest parties, Alice and Bob, who aim to generate a shared key that remains secret from any eavesdroppers, commonly referred to as Eve. By exchanging quantum signals, followed by classical postprocessing, QKD guarantees the security of the information against quantum adversaries with potentially infinitely many quantum resources \cite{renner2006security,Renner2022Security}.
In this regard, continuous-variable QKD (CVQKD) \cite{Diamanti2005Review,pirandola2020advances,zhang2024continuousvariable,Anka_2026}  allows the implementation of QKD protocols using coherent states \cite{grosshans2002continuous,grosshans_quantum_2003} and commercial telecommunication technologies. Hence, it provides affordability and network scalability \cite{LMariani2024}, especially over metropolitan distances where its high repetition rates can outperform discrete-variable implementations. 
Among the different CVQKD approaches, discrete-modulated (DM) protocols \cite{Leverrier2009TheoreticalSO,Leverrier2009RepetitionCode,Ghorai2019,Roumestan_2024}  have attracted special interest. In these schemes, the information is encoded in a finite constellation of coherent states, and the measurement outcomes are discretized into a reduced alphabet, thereby simplifying the postprocessing steps \cite{Leverrier2009RepetitionCode}.

Recent works have shown that DM CVQKD guarantees general security in the finite-size regime \cite{baeuml2023security,pascualgarcia2024,primaatmaja2024, navarro2025,bodenstedt2026}. However, they face important theoretical and practical challenges. 
A primary concern is that optical implementations underlying CVQKD protocols naturally operate in infinite-dimensional Hilbert spaces, complicating the numerical estimation of secret key rates. This challenge can be addressed by applying a cutoff assumption \cite{Lin2019Asymptotic, Ghorai2019}, which assumes a finite-dimensional representation for coherent states, albeit at the cost of making a mathematical assumption. As an alternative, dimension reduction was proposed, which projects the problem onto a finite-dimensional subspace at the cost of a penalty in the secret key rate \cite{Upadhyaya2021Cutoff}. Although it was shown that the removal of a photon-number cutoff is compatible with a general security framework thanks to the generalized entropy accumulation theorem (GEAT) \cite{dupuis2016entropy,dupuis2019entropy,metger2022security}, the cost of the penalization has proven to be a major hurdle in the practical analysis and implementation of DM CVQKD \cite{primaatmaja2024}. Specifically, this is caused by the need for a so-called affine min-tradeoff function for the GEAT, which induces a severe cost in the final secret key rate. 
 
Within this context, advances in security proofs, such as the marginal-constrained entropy accumulation theorem (MEAT) \cite{arqand2025MEAT}, have improved the GEAT framework to overcome the need for sequential structures, as well as the requirement of affine min-tradeoff functions. This approach relies on Rényi entropies to enhance the secret key generation rate \cite{Dupuis23RenyiHashLemma, kamin25MEATsecurity}, which can be efficiently computed thanks to new progresses in non-symmetric conic optimization  \cite{skajaa2015,papp2017,lorente2024,navarro2025,bodenstedt2026}, together with the facial reduction technique \cite{drusvyatskiy2017,hu2021robust}, for a fast, reliable computation of Rényi entropies \cite{git,git_examles}.

In this work, we develop a variable-length security analysis for DM CVQKD based on the MEAT \cite{arqand2025MEAT}, which surpasses the limitations of previous approaches in terms of repetition-rate restrictions, secret key rates, and cutoff assumptions. While the MEAT has previously been considered in the context of DM CVQKD under simplified assumptions \cite{navarro2025}, the present work extends this framework to a more realistic setting. To this end, we consistently integrate the dimension reduction technique \cite{Upadhyaya2021Cutoff} within the MEAT in order to accurately estimate numerical secret key rates without a cutoff assumption \cite{Lin2019Asymptotic, kanitschar22differentModulation}. We further complement our model for practical experimental conditions by incorporating a trusted detector model that accounts for receiver imperfections at Bob’s detectors \cite{LinTrusted2020}, alongside an enhanced postselection \cite{Lin2019Asymptotic,Kanitschar2023} that improves both test and the key distillation. To overcome the computational cost associated with the resulting high-dimensional Rényi optimization, we further introduce a hyperplane approximation of the Rényi cone. This approach provides efficiently computable and certifiable lower bounds on the secret key rate while allowing substantially larger dimension-reduction cutoff.
In particular, we focus our analysis on the fully discretized version of the quadrature phase shift keying (QPSK) protocol \cite{Leverrier2009RepetitionCode, baeuml2023security, pascualgarcia2024}, which offers an optimal balance between secret key rates and simplified postprocessing, while noting that the approach can be generalized to any other discrete modulations.

The rest of the article is structured as follows: in Section \ref{Sec:Preliminaries}, we provide the mathematical concepts used throughout this work. In Section \ref{Sec:ProtDescription} we briefly describe the protocol studied under our analysis, as well as the relevant definitions from QKD. Next, Section \ref{Sec:NumAnalysis}  presents the numerical model and results for our protocol, which are later discussed in Section \ref{Sec:Discussion}.

\section{Mathematical preliminaries} \label{Sec:Preliminaries}

Let us denote a state as \textit{classical-quantum} (cq-state) $\rho_{XY}$ when it can be expressed as 
\begin{equation}
    \rho_{XY} = \sum_{x\in \mathcal{X}} p(x) \ketbra{x}{x}_X\otimes \rho^x_Y\,,
\end{equation}
for some countable alphabet $\mathcal{X}$. Any subset $\Omega \subseteq \mathcal{X} $ can be used to define an event, which also suggests a restriction for cq-states according to a conditioning. This is expressed as

\begin{equation} \label{eq:TrPr}
    \rho_{XY|\Omega} = \frac{1}{\mathrm{Pr}_\rho [\Omega]}\sum_{x\in \Omega} p(x) \ketbra{x}{x}_X\otimes \rho^x_Y,
\end{equation}
with $\mathrm{Pr}_\rho[\Omega] = \sum_{x\in \Omega} p(x)$ representing the probability of observing the event. To simplify the notation, we will generally omit the subscript $\rho$ unless necessary for further clarification, and directly refer to $\Omega$ as the event. Let us now define the set of all subnormalized states in a Hilbert space $\mathcal{H}$ as $\mathcal{D}_{\leq}(\mathcal{H})$. For any two such states $\rho$ and $\sigma$, we define the generalized trace distance as

\begin{equation} \label{eq:GenTraceDist}
    T(\rho,\sigma) = \frac{1}{2} \lVert \rho - \sigma \rVert_1 +\frac{1}{2}\left|\Tr[\rho-\sigma]\right|.
\end{equation}
In particular, we observe that
\begin{equation} \label{eq:Trace2Prob}
    \mathrm{Pr}_\sigma[\Omega] = T(\sigma,\sigma_{|\neg\Omega}).
\end{equation}
These concepts become crucially important in QKD, especially when it is required to define the secrecy of a protocol as we will explain in the next sections. On the other hand, we will make use of diverse entropic quantities to define the secret information shared by Alice and Bob in relation to Eve -- the main one being the sandwiched, conditional Rényi entropy
\begin{equation}
    H^\uparrow_{\alpha}(A|B)_\rho = \sup_{\sigma_B \in \mc{D}(\mc{H}_B)} - D_\alpha(\rho_{AB}\|\mathds{1}\otimes\sigma_B)\,.
\end{equation}
Here, the right hand side is given by the sandwiched Rényi relative entropy
\begin{equation} \label{eq:QRenyiDiv}
    D_\alpha(\rho\|\sigma) = \frac1{\alpha-1}\log\left[\frac{\Psi_\alpha(\rho, \sigma)}{\Tr[\rho]}\right],
\end{equation}
where 
\begin{equation}\label{eq;PsiRenyi}
    \Psi_\alpha(\rho, \sigma) = \Tr\left[(\sigma^\frac{1-\alpha}{2\alpha}\rho\sigma^\frac{1-\alpha}{2\alpha})^\alpha\right],
\end{equation}
provided that $\alpha \in (0,\infty)\backslash\{1\}$ and $\mathrm{supp}(\rho) \subseteq \mathrm{supp}(\sigma)$. However, it will prove to be more useful to consider a lower bound, given by its down-arrow version
\begin{align} \label{eq:DownArrowRenyi}
    H^\downarrow_\alpha(A|B)_\rho := - D_\alpha (\rho_{AB}\|\mathds{1}_A \otimes \rho_B)\,.
\end{align}
This quantity can be used to define a more specialized family of entropies. Let now $\rho_{ABC}$ a state classical in $C$, whose support is the alphabet $\mc{C}$. Provided an arbitrary $f : \mc{C} \rightarrow \mathbb{R}$ denominated tradeoff function, the $f$-weighted Rényi entropy of order $\alpha  \in (0,\infty)\backslash\{1\}$ corresponding to the $\rho_{ABC}$ is \cite{arqand2024generalized,arqand2025MEAT}
\begin{equation}
     H^{\uparrow,f}_\alpha (A|BC)_{\rho}= \frac{\alpha}{1-\alpha} \log \left( \sum_{{c}\in \mc{C}} p({c}) 2^{\frac{1-\alpha}{\alpha} (-f ({c})+ H^\uparrow_\alpha (A|B))_{{\rho}_{|{c}}}}  \right).
\end{equation}
Another important entropic quantity that will be relevant in our discussion is the Kullback-Leibler divergence, denoted for two probability distributions $q,p \in \mathbb{P}_{\mc{X}}$ as
\begin{equation}
    D_{\mathrm{KL}}(q||p)= \sum_{x\in \mc{X}} q(x) \log \left( \frac{q(x)}{p(x)}\right)\,,
\end{equation}
where a version for binary probability distributions can be defined as
    \begin{equation} 
    d_\text{KL}(a||b) = a \ln \left(\frac{a}{b} \right) + (1-a)\ln \left( \frac{1-a}{1-b} \right).
\end{equation}
On the other hand, we denote a collection of $n$ identical registers as $C_1^n = C_1... C_n$, whose embedding is $C$. In the case that $C$ is classical, we denote its alphabet as $\mc{C}$ and the finite, frequency distribution for a particular combination $c_1^n \in \mc{C}^n$ as $\mathrm{F}^\mathrm{obs}_{c_1^n}$, with entries

\begin{equation}\label{eq:frequency distr.}
\text{F}^\text{obs}_{c_1^n}(\hat{c})=\frac{|\{j \in \{1, 2, \ldots, n\}
: c_j=\hat{c}\}|}{n}.
\end{equation}

\section{Protocol}\label{Sec:ProtDescription}

Let us provide a particular description of a CVQKD protocol based on the QPSK scheme, using a complete discretization of all the measurements \cite{Lin2019Asymptotic}. Our protocol is valid under the MEAT \cite{arqand2025MEAT} and reads

\begin{enumerate}
    \item \textit{Preparation and measurement}. For each round $j \in \{1,..., n \}$, Alice and Bob perform the following steps:
    \begin{itemize}
        \item Alice prepares one of the states $\{\ket{i^x {\gamma}}\}_{x=0}^3$ according to a uniform probability distribution and sends it to Bob. Alice records her input $x$. 
        \item Bob performs a heterodyne measurement followed by a discretization of his outcome according to a defined modulation and a binary random variable $Y_j \in \{\top,\bot\}$, with probabilities $\pkey$ and $1-\pkey$, respectively. If $Y_j= \bot$, the round is used for testing: Bob records his discretized measurement outcome in register $\tilde{Z}_j$ and sets $Z_j = \perp$. If $Y_j= \top$, the round is used for key generation: Bob records his discretized measurement outcome in the key register $Z_j$ and sets $\tilde{Z}_j = \bot$, or he sets  $Z_j = \perp$ if the signal is discarded due to a predefined postselection scheme. In the latter case, the discarded measurement outcome will be recorded on $\tilde{Z}_j$. 

        \item For each round, Bob constructs a register $I_j$ indicating whether his measurement is used for key distillation or not, and discloses $I \tilde{Z}_j$. Then, Alice stores her input $x_j$ in the registers $X_j$ or $\tilde{X}_j$, according to the value\footnote{In particular, we note that $\tilde{X}_j$ deterministically provides all information about $I_j$, such that the latter can be removed from our analysis.} of $I_j$, and discloses $\tilde{X}_j$.
    \end{itemize}
    
    \item \textit{Variable-length decision.} Using the observed announcement string $c_1^n$, Alice and Bob compute the frequency distribution $\mathrm{F}^\mathrm{obs}_{c_1^n}$ defined in Eq.~\eqref{eq:frequency distr.}. Then, they determine the final secret-key length $\ell(c_1^n)$ and the total number of bits $\mathrm{leak}(c_1^n)$ disclosed to perform information reconciliation.
    \item \textit{Information reconciliation.} Alice and Bob employ a reverse error correction scheme followed by error verification to eliminate the disagreements between their private keys. 
    \begin{itemize}
        \item Bob sends Alice $\lambda_{\mathrm{EC}}(c_1^n)$ bits of error correction information obtained from his raw key, which she uses to derive a guess of Bob's key.
        \item Alice independently samples a hash function $f(\cdot)$ from a universal$_2$ family, applies it to her guess, and sends to Bob both $f(\cdot)$ and the hashed key. 
        \item Bob applies $f(\cdot)$ to his private key and compares it to Alice's. If the hashed keys do not coincide, they abort.
    \end{itemize}

    \item \textit{Privacy amplification.} Following successful error verification, Alice and Bob distill the final secret key of length $\ell(c_1^n)$ via a privacy amplification method and applying an independent universal$_2$ hash function. 

\end{enumerate}

\begin{figure*}
    \begin{minipage}{.5\textwidth}
        \subfloat[]{\includegraphics[width=\textwidth]{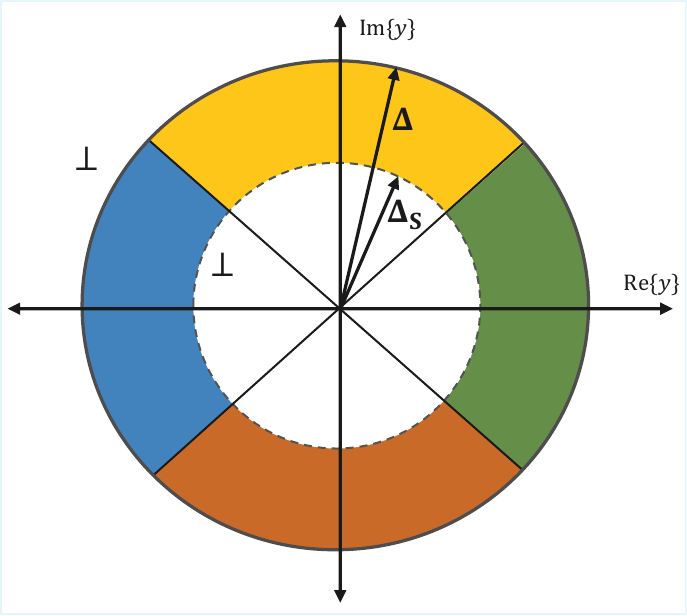}
        \label{fig:keyrounds}}
    \end{minipage}
    \hfill    
    \begin{minipage}{.5\textwidth}
        \subfloat[]{\includegraphics[width=\textwidth]{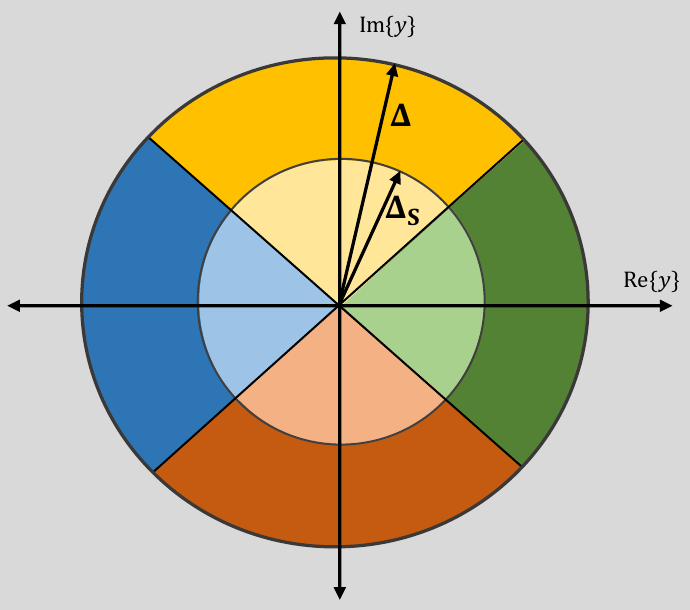}        
        \label{fig:parameterestimation}}
    \end{minipage}
        \caption{Phase space representation of the modulations used by Bob for (a) key distillation and  
        (b) test, according to the modulation parameters $\Delta_s$ and $\Delta$.}\label{fig:modulation}
    \end{figure*}

Based on this description, we now present a quantitative formulation of our protocol: for a round $j \in \{1,..., n \}$ with $n\in \mathbb{N}$, Bob's measurement outcome reads $y_j = |y_j|e^{i \theta_j}$, which is discretized for key rounds (i.e. $Y_j = \top$) using a binning that divides the phase space into five regions, as shown in Figure \ref{fig:keyrounds}. He stores his discretized measurement outcome in the following register:
\begin{equation}\label{eq:discKey}
{Z}_j=\begin{cases}
z &\text{ if }{ \theta_j\in[(2z-1)\pi/4,(2z+1)\pi/4)} \wedge|{y}_j|\in (\Delta_s,\Delta) \wedge Y_j = \top,\\
\perp &\text{ else, }
\end{cases}
\end{equation}
where $z\in \{0,...,3\}$. For later reference, we define the alphabet of key values ${\mc{Z}}=\{0,...,3,\perp\}$, and $\hat{\mc{Z}}= \{0,...,3\}$ after removing the symbol for discarded rounds. In particular, this binning directly discards rounds when $|y_j| \notin (\Delta_s,\Delta)$, according to the modulation parameters $\Delta_s > 0$ and $\Delta > \Delta_s$, which we can use in order to define the postselection register
\begin{equation}
    I_j = \begin{cases}
        \top & \text{ if } {Z}_j \neq \perp \,,\\
        \bot & \text{ else. }
    \end{cases}
\end{equation}
For test rounds, we may use the parameters $\Delta_s$ and $ \Delta$ following a more involved discretization as shown in Figure \ref{fig:parameterestimation}, such as

\begin{equation}\label{eq:discPE}
\tilde{Z}_j=\begin{cases}
z &\text{ if } {{\theta}_j\in[(2z-1)\pi/4,(2z+1)\pi/4)\land|{y}_j|\in [0,\Delta_s)}, \\
4+z &\text{ if }{{\theta}_j\in[(2z-1)\pi/4,(2z+1)\pi/4)\land|{y}_j| \in [\Delta_s,\Delta)} \wedge  I_j = \bot,\\
8 &\text{ if } |{y}_j| \geq \Delta, \\
\perp &\text{ else,}
\end{cases}
\end{equation}
again for $z\in \{0,...,3\}$. Once Bob sends $[I \tilde{Z}]_j$ to Alice, she stores her input $x_j$ in two new registers

\begin{align} \label{eq:AliceKey}
{X}_j&=\begin{cases}
x_j & \text{ if }I_j= \top, \\ % \wedge Y_j = \top,\\
\perp &\text{ else}.
\end{cases}\\
\tilde{X}_j&=\begin{cases}\label{eq:AlicePE}
x_j & \text{ if }I_j= \bot, \\ % \vee  Y_j = \bot,\\
\perp & \text{ else},
\end{cases}
\end{align}
such that the formulation explicitly splits the rounds ${X}_j$ and $\tilde{X}_j$ for key generation and test. In order to simplify the notation, we merge the registers for test as $\tilde{C}_j := [\tilde{X} \tilde{Z}]_j$, given by an alphabet 
\begin{equation}
    \mc{C} = (\bot, \bot) \cup (\{0,...,3\} \times  \{0,...,8\}). % \cup (\{\bot\} \times \{0,...,3\} \times  \{0,1,2,3,8\}).
\end{equation}
Furthermore, we define the alphabet  $\tilde{\mc{C}} := \mc{C} \backslash (\bot, \bot )$ for nontrivial test values. After a successful test, Alice and Bob proceed to perform error correction via a pre-defined algorithm (e.g. LDPC codes or turbocodes) which leaks a maximum amount of bits to the adversary according to a pre-defined value. After this step, Alice has a guess $\hat{X}_j$ of every bit at Bob's secret key ${Z}_j$, and employs a universal$_2$ hash function $f(\cdot)$ on her guess, resulting in a hash\footnote{With a slight abuse of notation, let us consider that the hash function also removes the $\{\perp\}$ symbols from the raw keys.} $H = f(\hat{X}_1^n)$ of length $\lceil \log(1/\varepsilon_\mathrm{cr}) \rceil$ according to a previously agreed coefficient $\varepsilon_\mathrm{cr} \in (0,1)$. Similarly, we define Bob's hash as  $H' = f({Z}_1^n)$, which he compares with Alice's in order to verify the success of the error correction. If they coincide, Alice and Bob apply a different hash to perform privacy amplification, or abort otherwise. For simplicity, we defer the definition of Bob's measurement, defined by positive-operator valued measures (POVMs) $\{R^z_B \}_{z \in \mc{Z}}$, and $\{\tilde{R}^z_B \}_{z=0}^8$, to Section~\ref{Sec:NumAnalysis}.

\subsection{Security for QKD}\label{Subsec:Security}

We define the notion of security for QKD within the well-known $\varepsilon-$security framework developed by Renner \cite{renner2006security}. According to this approach, Alice and Bob aim to achieve a target ccq-state shared with Eve, represented as

\begin{align}\label{eq:IdealCCQ}
    \bar{\tau}_{K_A K_B E} &= \frac{1}{2^\ell}\sum_{x=0}^{2^\ell-1} \pro{x,x}_{K_A K_B}\otimes \rho_E = \tau_{K_A K_B} \otimes \rho_E.
\end{align}
In this ideal configuration, Alice and Bob share perfect classical randomness in their respective registers $K_A$ and $K_B$, which remain completely uncorrelated with respect to Eve's system, in state $\rho_E$. As a result, their registers can be used as a secret key of $\ell$ bits. However, in a realistic scenario, experimental imperfections and potential attacks from Eve make this state unreachable. Instead, Alice and Bob's objective will be to achieve a state that is sufficiently close to \eqref{eq:IdealCCQ}, and abort the protocol otherwise. This idea is formalized according to the indistinguishability between the ideal and real cases up to a margin $\epsilon_s$, such that a state $\rho_{K_A K_B E}$ is considered $\varepsilon_\mathrm{s}-$secure if it satisfies\footnote{The notion of $\varepsilon-$security actually allows Eve may hold different marginals between the ideal and real states \cite{ferradini2025definingsecurityquantumkey}. This comes at the reduced cost of including an extra 2 factor.}
\begin{equation} \label{eq:secure}
    T( \rho_{K_A K_BE}, \bar{\tau}_{K_A K_BE}) \leq \varepsilon_\mathrm{s}.
\end{equation}
Within a variable-length framework \cite{Tupkary2024}, the security statement \eqref{eq:secure} is generalized to
\begin{equation} \label{eq:secureVarLen}
    \sum_\ell \mathrm{Pr}_\rho[\Omega_\ell]
    T( \rho_{K_A K_BE|\Omega_\ell},\bar{\tau}_{K_A K_BE|\Omega_\ell} )\leq \varepsilon_\mathrm{s},
\end{equation}
where we decompose the initial formula according to all possible key lengths $\ell$ that Alice and Bob can generate, weighted by their corresponding probabilities $\mathrm{Pr}_\rho[\Omega_\ell]$, with $\Omega_\ell$ denoting that the protocol outputs keys of length $\ell$ after successful verification. In particular, an implementation such that $\ell = 0$ trivially leads to the ideal state, without generating any secret key.

The definition for security can be simplified according to the clauses of correctness and secrecy. A QKD protocol is $\varepsilon_\mathrm{EV}-$correct when Alice's and Bob's keys disagree without aborting with a probability not larger than $\varepsilon_\mathrm{EV}$. Mathematically, this is expressed as

\begin{equation}\label{eq:correctness}
    \sum_\ell\mathrm{Pr}_\rho [(K_A \neq K_B) \wedge \Omega_\ell] \leq \varepsilon_\mathrm{EV}.
\end{equation}
This condition is satisfied via the validation hash $f(\cdot)$ at the end of information reconciliation \cite{Tupkary2024,kamin25MEATsecurity,navarro2025}. On the other hand, an $\varepsilon_\mathrm{PA}-$secret implementation ensures that Eve is $\varepsilon_\mathrm{PA}-$close to being fully uncorrelated with respect to one of the keys

\begin{equation} \label{eq:secret}
    \sum_\ell \mathrm{Pr}_\rho[\Omega_\ell]
    T( \rho_{K_BE|\Omega_\ell},\sigma_{K_BE|\Omega_\ell} )\leq \varepsilon_\mathrm{PA},
\end{equation}
where $\sigma_{K_BE|\Omega_\ell} :=\frac{1}{2^\ell} \mathds{1}_{K_B}\otimes \rho_{E|\Omega_\ell}$. Secrecy can be enforced via the privacy amplification hash, which leverages the so-called leftover hash lemma \cite{tomamichel2015quantum,berta2016smooth}. In our case, we will use a version that employs Rényi entropies \cite{Dupuis23RenyiHashLemma}, which can be later used to quantify the secret key length via the MEAT (see Appendix \ref{App:PrivacyAmp} for a further discussion). Once a protocol verifies $\varepsilon_\mathrm{EV}-$correctness and $\varepsilon_\mathrm{PA}-$secrecy, it can be verified via a triangle inequality on \eqref{eq:secure} that it is $(\varepsilon_\mathrm{EV}+\varepsilon_\mathrm{PA})-$secure (see for instance Ref. \cite{Kanitschar2023}).

\subsection{Channel description}\label{Subsec:ChannelDescription}
In order to quantitatively derive the secret key rate, it is necessary to describe the full procedure for one round given the actions of Alice and Bob. To this end, we switch from the prepare-and-measure image into its entanglement version, which is fully equivalent thanks to the source-replacement scheme \cite{BBM92}. In said image, Alice actually prepares entangled states and performs measurements in a computational basis . Provided the description of the protocol and denoting the quantum state shared by Alice and Bob via registers $AB$, we represent the one-round channel $\mc{M}: AB \to ZC$ as
\begin{align}
    \mc{M}(\cdot)_{ZC} &= \bar{p}(\perp) \sum_{z \in \hat{\mc{Z}}} \Tr[\id_A \otimes R^z_B (\cdot)] \pro{z}_Z \otimes \pro{\perp}_C  \nonumber\\
    & +    \sum_{(x,z) \in \tilde{\mc{C}}} \bar{p}(x,z)\Tr[\pro{x}_A \otimes \tilde{R}^z_B (\cdot)]\pro{\perp}_Z \otimes \pro{x, z}_C, 
    \label{eq:CompleteMap}
\end{align}
where we define the masking vector $\bar{p}(c)$ as

\begin{align}\label{eq:PselMask}
    \bar{p}(c) = \begin{cases}
        \pkey, &\, \text{ if } \, c = \perp, \\
        (1-\pkey),& \, \text{ if } c = (x,z) \text{ with } x \in \{0,1,2,3\} \wedge z \in \{4,5,6,7\}, \\
        1, &\, \text{else}. \\
    \end{cases}
\end{align}
On the other hand, let us also consider a generalization of these maps whenever we take the input state with a purifying register $E$. We perform this extension with a tilde, such as 
\begin{equation}\label{eq:ExtendedMap}
     \mmeas(\cdot)_{EZC} = \id_{E} \otimes \mc{M}(\cdot)_{ZC},
\end{equation}
for the one-round map.

\subsection{Dimension reduction}
As a final step in our protocol outline, it is important to recall that CVQKD fundamentally handles infinite-dimensional states. This becomes a major hurdle in EAT-based security proofs \cite{pascualgarcia2024,primaatmaja2024}, which require a numerical optimization over quantum states. 
Hence, DM CVQKD is typically modeled via a cutoff \cite{Lin2019Asymptotic,lupo_quantum_2022, baeuml2023security, staffieri2025renyidvcvqkd,navarro2025}, such that the dimension of Bob's register is limited as $\dim(\mc{H}_B) = N_c +1$. Although this mathematical simplification allows a practical calculation of secret key rates, neglecting contributions outside the cutoff without
rigorously bounding their effect constitutes a security gap.

Alternatively, it is possible to use a dimension reduction \cite{Upadhyaya2021Cutoff}, such that Bob's register is projected onto a finite-dimensional subspace of dimension $N_c + 1$, whereas the contributions from the complementary subspace and the cross terms between the two subpaces is bounded. For instance, choosing the subspace spanned by the Fock states $\{\ket{n}\}_{n=0}^{N_c}$ allows an efficient characterization of the weight $\Tr[\Pi_B \omega]$ of any state $\omega \in \mc{D}(\mc{H}_B)$. This is compatible with general security proofs, as Bob's POVM can eventually be used to bound said weight. In particular, Ref. \cite{primaatmaja2024} applied this approach under the generalized EAT \cite{metger2022security}, albeit leading to excessive penalties due to the limitations of the security proof. 

In Appendix \ref{app:DimRed}, we apply this approach to finite-dimensional
approximations of arbitrarily large size $m>N_c$, keeping the numerical cutoff $N_c$ fixed and deriving bounds independent of $m$. In particular, we use continuity bounds for the sandwiched Rényi entropies \cite{Marwah_2022, Bluhm_2024} that introduce a calculable penalty to the secret key rate. Together, these bounds lead to the finite-dimensional optimization in Appendix \ref{App:ConicKappa}, which provides a secure key-length bound valid for every $m>N_c$. Note that the resulting security guarantee is extended to the physical
infinite-dimensional protocol by trace-norm approximation, as
shown in Appendix \ref{app:SecurityProof}.

\subsection{Finite-size general security} \label{sec:finite size security}

Provided the previous discussion and definitions, we can define the security and practical output of the protocol according to the following results.
\begin{theorem}[Variable-length secret key]\label{Th:MainTheorem}
    Let  $n, N_c \in \mathbb{N}$, $\sigma_A \in \mc{D}(A)$ and ${\mc{M}}:[AB]_j \to [ZC]_j$ a quantum channel concatenated $n$ times to form the protocol. Let further $C$ be a classical register defined on an alphabet $\mc{C}$, and $\Pi_B = \sum_{j=0}^{N_c}\pro{j}_B$ a finite-dimensional projector. Given a set of non-negative parameters $\varepsilon_\mathrm{EV},\varepsilon_\mathrm{PA}$, the protocol is $\varepsilon_\mathrm{EV}$-correct and $\varepsilon_\mathrm{PA}$-secret if the secret key length is set according to
    \begin{align} \label{eq: final key}
    \ell \leq \max \left\{0, n\bigl(f\cdot \mathrm{F}^\mathrm{obs}_{c_1^n}+\kappa\bigr)
    -\mathrm{leak}(c_1^n)
    -\frac{\alpha}{\alpha-1}
    \log \left( \frac1{\varepsilon_{\mathrm{PA}}} \right)
    +2\right\},
    \end{align}
where $\alpha \in (1,2)$, $f:\mc{C} \to \mathbb{R}$ a tradeoff function (optimally chosen from a reference probability distribution  $p \in \mathbb{P}_{\mc{C}}$), $\mathrm{F}^\mathrm{obs}_{c_1^n}$ a frequency distribution defined in Eq. \eqref{eq:frequency distr.}, $\mathrm{leak}(c_1^n)$ the information lost to Eve during an $\varepsilon_\mathrm{EV}$-correct information reconciliation calculated according to the test registers $C_1^n = c_1^n$; and $\kappa$ a scalar resulting from a conic optimization (see Appendix \ref{app:FixedLength} and the formulation in Eq. \eqref{eq:KappaConic}) given by a the channel model $\mc{M}$ and its corresponding key distillation maps $\g$ and $\zg$.

\end{theorem}

The proof for this statement can be found in Appendix \ref{app:SecurityProof}.

\subsection{Information Reconciliation}

We characterize the error correction cost by bounding the number of bits that Alice and Bob exchange over the public channel to reconcile their private keys. We assume one-way reverse reconciliation on the non-postselected rounds. In the ideal scenario, the Slepian-Wolf theorem \cite{SlepianWolf} dictates that the minimum amount of information Bob must send to Alice is $H(Z|X) = H(Z) - I(X:Z)$, where $I(X:Z)$ is the mutual information between Alice's and Bob's registers. In practice, however, the reconciliation protocol is not perfectly efficient, and we therefore need to introduce a reconciliation efficiency parameter $\beta \in [0,1]$, where $\beta = 1$ represents the Shannon limit. Consequently, the practical information leakage per reconciled symbol is expressed as
 \begin{equation}
      H(Z)-\beta I(X:Z)=(1-\beta)H(Z) + \beta H(Z|X).
 \end{equation}
Considering the total number of exchanged signals, the total leakage during information reconciliation  is
bounded by
\begin{align}\label{eq:EC_P2P}
    \mathrm{leak}(c_1^n) \leq  n \lambda_{\mathrm{EC}}(c_1^n) + \left\lceil \log \left(\frac{1}{\varepsilon_\mathrm{EV}} \right) \right\rceil .
\end{align}
Here, $\lambda_{\mathrm{EC}}(c_1^n)= \mathrm{F}^\mathrm{obs}_{c_1^n}(\perp) [(1-\beta)H(Z) + \beta H(Z|X)] $ denotes the prescribed
length, in bits, of Bob's error-correction message for the
observed announcement string $c_1^n$, where the pre-factor $\mathrm{F}^\mathrm{obs}_{c_1^n}(\perp)$ indicates the fraction of the total amount of rounds that are not postselected, and thus require to be error-corrected. The term $\left\lceil \log \left(1/\varepsilon_\mathrm{EV} \right) \right\rceil$ represents the bits lost to the adversary during the validation of the error-corrected keys.

\section{Computing the key rate} \label{Sec:NumAnalysis}
\subsection{Alice's marginal}
Now, we are in a position to provide a numerical characterization for the secret key length introduced in Theorem \ref{Th:MainTheorem} by using the entanglement-based picture again \cite{E91} for simplicity. In this image, Alice always prepares the same state $\ket{\psi}$ for all rounds according to an amplitude $\gamma \in \mathbb{R}$, such that
\begin{align}\label{eq:InitialState}
\ket{\psi} = \frac{1}{2}\sum_{j=0}^3\ket{j}_A \otimes \ket{i^j \gamma}_{A'}.
\end{align} 
Consequently, Alice's marginal $\sigma_A$ results in

\begin{equation}
\sigma_A = \Tr_{A'}[\pro{\psi}_{AA'}] = \frac{1}{4}\sum_{j,k=0}^3 \braket{i^k \gamma}{i^j \gamma} \ket{j}\bra{k}_A.
\end{equation}

\subsection{Noisy region operators and probabilities} \label{sec:operators and prob}
While Alice's measurements, in an entanglement-based picture, correspond to projective measurements in her computational basis, Bob's are provided by a coarse-grained heterodyne measurement. We characterize Bob's measurement according to a trusted, non-ideal detector model presented in \cite{LinTrusted2020}, assuming finite detection efficiency $\eta_d$ and some electronic noise $\nu$ in the detectors of the homodyne measurements composing the heterodyne detection.\footnote{The detection efficiency and electronic noise may in general differ for both homodyne detectors, but for simplicity we take them to be equal.}

These imperfections are described by placing a beamsplitter of transmissivity $\eta_d$ before the homodyne detection, and then mixing the incoming signal with a thermal state of mean photon number $\bar{n} = (1-\eta_d + \nu)/\eta_d$. As a consequence, Bob’s heterodyne measurement is no longer described by projections onto pure coherent states, but rather by projections onto displaced thermal states incorporating the effects of detector inefficiency and electronic noise. The resulting POVM can therefore be written as (see \cite{LinTrusted2020} for the complete derivation)
\begin{equation}
    R (\gamma e^{i \theta},\eta_d,\nu) = \frac{1}{\eta_d \pi} \bar{D}\left(\frac{\gamma e^{i \theta}}{\sqrt{\eta_d}} \right) \rho_\mathrm{th}\left(\frac{1-\eta_d + \nu}{\eta_d} \right) \bar{D}^\dagger\left(\frac{\gamma e^{i \theta}}{\sqrt{\eta_d}}\right).
\end{equation}
where $\bar{D}(\cdot)$ is the displacement operator. Therefore, for rounds spent on the statistical checks of test, the POVM reads
\begin{subequations}\label{eq:OperatorsPE}
\begin{align}
     \tilde{R}^z_B &= \int_{0}^{\Delta_s} \int_{\frac{\pi}{4} (2z-1)}^{\frac{\pi}{4}(2z+1)}  R (\gamma e^{i \theta},\eta_d,\nu)\gamma \;  d\theta d\gamma,  && z \in \{0,1,2,3\},  \\
     \tilde{R}^z_B &= \int_{\Delta_s}^\Delta \int_{\frac{\pi}{4} (2z-1)}^{\frac{\pi}{4}(2z+1)}  R (\gamma e^{i \theta},\eta_d,\nu) \gamma \;  d\theta d\gamma,  && z \in \{4,5,6,7\},  \\
     \tilde{R}^8_B &= \int_\Delta^\infty \int_0^{2 \pi} R (\gamma e^{i \theta},\eta_d,\nu) \gamma \;   d\theta d\gamma.
\end{align}
\end{subequations}
In particular, the non-trivial key measurements are given by the subset $\{\tilde{R}^z\}_{z=4}^7$, which we will identify as $\{{R}^z\}_{z \in \hat{\mc{Z}}}$. In consequence, the postselected region is given by $R^\bot_B = 1-R^\top_B$, where we identify $R^\top_B$ as $ \sum_{z\in\hat{\mc{Z}} } R^z_B$.

On the other hand, the reference distribution $p^\mathrm{r}$ is taken by simulating an honest implementation, where the measurements of Bob are provided by the discretization introduced in Section \ref{Sec:ProtDescription}, as well as Alice's chosen state \eqref{eq:InitialState}. Regarding the quantum channel, we model it as an optical fiber characterized by its excess noise $\xi$ and efficiency $\eta = \eta_d 10^{-\chi/10}$, where $\chi$ represents the total channel transmittance in dB. To ease the notation, we define \cite{LinTrusted2020} 
\begin{equation}
    p(y e^{i \theta},x, \eta, \nu,\xi) = \frac{\upsilon}{4\pi} \exp\left(- \upsilon| ye^{i  \theta} - \sqrt{\eta}i^x \gamma  |^2\right),
\end{equation}
where $\upsilon = 1/(1 + \eta \xi/2 + \nu)$. Then, the simulated, reference probability distribution for test rounds is provided by 
\begin{equation}
    p(x,z) = \bar{p}(x,z) \tilde{p}(x,z)
\end{equation}
where we have
\begin{subequations}
\begin{align} 
     \tilde{p} (x,z) &=   \int_0^{\Delta_s} \int_{\frac{\pi}{4} (2z-1)}^{\frac{\pi}{4}(2z+1)}    p(y e^{i \theta},x, \eta, \nu,\xi) y \; dy  d\theta, & z \in \{0,1,2,3\},\label{eq:SiftingProb1} \\
     \tilde{p} (x,z) &=  \int_{\Delta_s}^\Delta \int_{\frac{\pi}{4} (2z-1)}^{\frac{\pi}{4}(2z+1)}   p(y e^{i \theta},x, \eta, \nu,\xi) y \; dy  d\theta, & z \in \{4,5,6,7\}. \label{eq:SiftingProb2} \\
     \tilde{p} (x,8) &=  \int_{\Delta}^\infty \int_{0}^{2\pi } p(y e^{i \theta},x, \eta, \nu,\xi)y \; dy d\theta, \label{eq:SiftingProb3}
\end{align}
\end{subequations}
with $x \in \{0,1,2,3\}$. From here, we can define the sifting probability $p_s = 1 - \sum_{x=0}^3\sum_{z = 0,1,2,3,8} \tilde{p}(x,z)$, i.e., the probability that the round is not discarded according to the postselection criterion. This also defines the simulated probability for $c= \perp$, such that
\begin{equation}
    p(\perp) = \bar{p}(\perp) p_s.
\end{equation}

\subsection{Conic formulation and hyperplane approximation}\label{sec:hyperplane}

As established in Section \ref{sec:finite size security}, the application of the dimension reduction technique successfully projects an arbitrary large infinite-dimensional system of dimension $m>N_c$ onto a smaller finite-dimensional subspace of dimension $N_c$ without compromising security. However, while the state space is finite, the resulting minimization of the secret key rate still requires specialized numerical methods to be solved efficiently.

To make this computation tractable, we frame the minimization as a convex optimization problem \cite{boyd2004convex,Nesterov2018}. Specifically, we apply the recently developed non-symmetric conic optimization methods introduced in \cite{navarro2025}. This approach allows us to express, in terms of exact conic representations, from the Rényi entropic quantities the corresponding physical constraints (such as Alice's marginal) to the penalization due to the dimension reduction and the observed test statistics. To further reduce the computational overhead, our methodology incorporates a facial reduction technique \cite{drusvyatskiy2017,hu2021robust} as presentend in Appendix \ref{app:Facial}, which greatly simplifies the numerical cost of the minimization \cite{navarro2025}. The complete reformulation of Theorem \ref{Th:MainTheorem} into the explicit conic program is detailed in Appendix \ref{App:VarLengthKeys}.

Despite these reductions, solving the conic program directly remains challenging. In practice, obtaining non-trivial secret key rates typically requires a cutoff of $N_c \geq 20$. Such a value leads to excessively large runtimes, and overall limited convergence. To address this difficulty, we bound the sandwiched Rényi entropy cone by a collection of hyperplanes. This produces an outer approximation of the original conic problem and a lower bound to the secret key rate. The approximation can be solved more efficiently because it requires only semidefinite constraints, inequality constraints, the second order cone, and the power cone. To construct a good approximation we need a near-optimal solution for the initial conic program, which can be readily found by computing the optimal solution for a smaller dimension $N'_c$, padding it with zeros, and depolarizing.

This approach has the added advantage that it allows us to compute the dual problem analytically (see Appendix \ref{sec:dual}). Since any point that satisfies the constraints of the dual problem gives us a lower bound for the primal problem, the dual solution allows us to certify lower bounds for the key rate without trusting the solver, as we can simply verify that the constraints are indeed satisfied. This is not possible with the sandwiched Rényi entropy cone, as no closed-form expression is known for its dual, and thus further optimization is needed to verify that a point belongs to the dual cone. We defer the quantitative description of this methodology to Appendix \ref{app:Hyperplanes}, and follow with an analysis of the results.

\subsection{Results}\label{subsec:Results}

Given the previous descriptions, we can now showcase the yield of our method by calculating the secret key rate for the QPSK protocol using conic programming \cite{navarro2025}. For the numerical evaluation, we compute the expected key rate and therefore use the honest distribution $p^h$ in place of the expected observed frequency vector $\mathds{E}[\mathrm{F}^\mathrm{obs}_{c_1^n}]$. In particular, $p^h(\perp)=p^{\mathrm{K}}p_s$, so that the expected number of reconciled symbols in Eq. \eqref{eq:EC_P2P} is $\mathds{E}[\mathrm{F}^\mathrm{obs}_{c_1^n}(\perp)]=p^{\mathrm{K}}p_s$. To model a non-ideal channel and measurements, we set the excess noise $\xi=0.005$ SNUs, and model the imperfections at the receiver via a detection efficiency $\eta_d = 0.71$ and electronic noise $\nu = 0.02$ SNUs, motivated by the experimental parameters reported in \cite{Hajomer2025experimentalDMCV}. Additionally, we fix the reconciliation efficiency to $\beta = 95\%$, and assume the following values for the tolerance coefficients \cite{arqand2024generalized}
\begin{align}
    \varepsilon_\mathrm{EV} = \frac{\alpha-1}{2 \alpha-1} \varepsilon  &&
    \varepsilon_\mathrm{PA} = \frac{\alpha}{2 \alpha-1} \varepsilon
\end{align}
which together define an $\varepsilon-$secure implementation of the protocol, setting $\varepsilon=10^{-10}$.

\begin{figure}[!ht]
	\centering
 	\begin{tikzpicture}
		\begin{axis}[%
            scale only axis,
            ymode = log,
            xmin=0, xmax=10,
            ymin=5e-4, ymax=1e-0,
            grid=major,
			xlabel={Transmittance $\chi$ (dB)},
            ylabel = {Secret key rate (bits/pulse)},
			axis background/.style={fill=white},
			legend style={at={(0.97,0.97)},legend cell align=left, align=left, draw=white!15!black, font=\footnotesize}
			]
            \addplot[orange, mark=*] table[col sep = comma, x index=0, y index = 1] {plot_data/hyperplanes/compare_n1e8.csv};  
            \addlegendentry{Hyperplanes, $(N_c,N_c') = (10,30)$}
            \addplot[red, mark=*] table[col sep = comma, x index=0, y index=3]{plot_data/hyperplanes/compare_n1e8.csv};
        	\addlegendentry{Hyperplanes, $(N_c,N'_c) = (16,30)$}
             \addplot[black!80, mark=*, dashed] table[col sep = comma, x index=0, y index=4]{plot_data/hyperplanes/compare_n1e8.csv};
        	 \addlegendentry{Hyperplanes, $(N_c,N'_c) = (16,45)$}
                 \addplot[teal, mark=*] table[col sep = comma, x index=0, y index=6]{plot_data/hyperplanes/compare_n1e8.csv};
        	\addlegendentry{Hyperplanes, $(N_c,N'_c) = (20,30)$}
            \addplot[olive, mark=*] table[col sep = comma, x index=0, y index=5]
            {plot_data/hyperplanes/compare_n1e8.csv};
            \addlegendentry{True cone, $\quad \; \, N_c=20$}
        \end{axis}
	\end{tikzpicture}
    \caption{Finite-size key rates against the total transmittance $\chi$ for $n= 10^{8}$ and different dimension reductions $N_c$ and $N_c'$. The curves under the approximation use three hyperplanes. We fixed $\Delta = \{2.5,3.\}$, and set a reconciliation efficiency $\beta=95\%$, while $\alpha$, $\gamma$, $\pkey$ and $\Delta_s$ were optimized for each data point.}\label{fig:benchmark hyperplane}
\end{figure} 

Figure \ref{fig:benchmark hyperplane} presents the finite-size secret key rates as a function of the total transmittance $\chi$ for both the conic method and the hyperplane approximation, considering different combinations of $N_c$ and $N_c'$ (see Section \ref{sec:hyperplane}). Here we observe that directly solving the conic program with $Nc=20$ leads to numerical instabilities and yields positive key rates only up to 9 dB. For the hyperplane approximation, increasing $N_c$ from 10 to 16 improves the achievable key rates, with additional gains by taking $Nc=20$. We nevertheless choose $N_c=16$ for the next plots as a compromise between key rate performance and computational cost. In particular, reducing the effective dimensionality lowers the runtime from approximately $1.6$ hours per point for $(N_c,N_c')=(20,30)$ to around $27$ min per point for $(N_c,N_c')=(16,30)$, while maintaining comparable key rates. This reduction also comes as an advantage on the memory needed, facilitating parallelization. Furthermore, the hyperplane approximation allows arbitrary values of $N_c'$, however, increasing $N_c'$ from $30$ to $45$ leaves the key rates almost unchanged, motivating our choice of $N_c'=30$. Such an increase is possible since it only affects the performance of the hyperplane programs (see Appendix \ref{app:Hyperplanes}), which offer a much better numerical scalability than the actual cone.

\begin{figure}[!ht]
	\centering
 	\begin{tikzpicture}
		\begin{axis}[%
			scale only axis,
            ymode = log,
			xmin=0,
			xmax=12,
			ymin=1e-4,
			ymax=1e-0,
			grid=major,
			xlabel={Transmittance $\chi$ (dB)},
            ylabel = {Secret key rate (bits/pulse)},
			axis background/.style={fill=white},
			legend style={at={(0.97,0.97)},legend cell align=left, align=left, draw=white!15!black, font=\footnotesize}
			]
            \addplot[black!70, mark=*] table[col sep = comma, x index=0, y index = 1] {plot_data/hyperplanes/Hyperplanes.csv};  
            \addlegendentry{$n = 10^{9}$}   
            \addplot[red, mark=*] table[col sep = comma, x index=0, y index=2]{plot_data/hyperplanes/Hyperplanes.csv};
            \addlegendentry{$n = 10^{8}$}
            \addplot[purple, mark=*] table[col sep = comma, x index=0, y index=3]{plot_data/hyperplanes/Hyperplanes.csv};
        	\addlegendentry{$n = 10^{7}$}
            \addplot[blue, mark=*] table[col sep = comma, x index=0, y index=4]{plot_data/hyperplanes/Hyperplanes.csv};
        	\addlegendentry{$n = 10^{6}$}
        \end{axis}
	\end{tikzpicture}
    \caption{Finite-size key rates against the total transmittance $\chi$ for different block sizes using the hyperplane approximation. All points employ a reduced dimension $N_c =16$, with at least three hyperplanes under an expanded dimension $N_c' = 30$; a modulation of $\Delta \in [2,4]$, and reconciliation efficiency $\beta=95\%$. The parameters $\alpha$, $\gamma$, $\pkey$ and $\Delta_s$ were optimized for each data point. } \label{fig:hyperplanes}
\end{figure}

Using the configuration $(N_c,N_c')=(16,30)$ motivated above, Figure \ref{fig:hyperplanes} presents the finite-size secret key rates for different block sizes $n$, computed via the dual problem in the hyperplane approximation (see Appendix \ref{sec:dual}). Block sizes similar to those obtained in this work have been reported in recent literature \cite{pascualgarcia2024, navarro2025}; however, these works rely on the cutoff assumption and do not consider trusted noise (see Table \ref{tab:qkd_comparison}). \footnote{We particularly note that \cite{baeuml2023security,pascualgarcia2024} use a scaling $f$ for the error correction efficiency, in line with the standard notation for discrete variable QKD.}  The latter assumes that the detector efficiency and electronic noise it is not accessible to Eve, leading to much higher key rates \cite{LinTrusted2020}. Moreover, the analysis in \cite{navarro2025} assumes an ideal error correction operating at the Shannon limit. Our work also differs from previous approaches in the security proof used to deal with general attacks. Specifically, we employ the MEAT framework, which avoids constructing an affine min-tradeoff function. Optimizing such functions can be technically challenging and may introduce additional looseness in finite-size analyses, resulting in more conservative secret key rate estimates. Furthermore, MEAT can be applied to prepare-and-measure protocols without requiring virtual tomography procedures or imposing sequential constraints on the implementation. Consequently, the protocol is not subject to the repetition-rate limitations associated with previous security-proof techniques (see \cite{arqand2025MEAT} for further details).

For the simulations, the parameters $\gamma$, $\Delta_s$, and  $\alpha$, were independently optimized for each transmittance point to maximize the secret key rate. While adapting the modulation parameter for each transmittance could yield further improvements, we fix $\Delta$ based on the block size and chosen dimension-reduction cutoff, since it showed a flat behavior in certain regions. Furthermore, we observe that the optimal value of $\pkey$ oscillates between 0.99 and 0.95 across the evaluated block sizes. Since any postselected signal is actively recycled for test, this indicates that our postselection scheme already accumulates sufficient statistics to tightly bound the channel parameters.  Consequently, Bob can maximize his raw key by allocating most of the incoming signals to key distillation, without needing to explicitly sacrifice a larger fraction for test.

\begin{table}[h!]
    \centering
    \caption{Comparison of DM CVQKD protocol parameters and security assumptions for diverse references.}
    \label{tab:qkd_comparison}
    \small 
    \begin{tabularx}{\textwidth}{@{} p{2cm} X c c c c X @{}}
        \toprule
        \textbf{Reference} & \textbf{Bound on\newline Dimension} & \textbf{$N_c$} &  \textbf{Block-Size $n$} & \textbf{Trusted Noise} & \textbf{Security  \newline Base} & \textbf{Efficiency \newline of EC } \\ 
        \midrule
        Upadhyaya, \newline et. al. \cite{Upadhyaya2021Cutoff} & Dimension reduction & 20& $\infty$ & Yes & Devetak- \newline Winter & $ 0.95$ \\ \addlinespace
        Kanitschar, \newline et. al. \cite{Kanitschar2023} &  Dimension reduction & 20  & $10^{9} - 10^{12}$ & Yes & Collective  \newline  \newline(AEP) & $0.95$\\ \addlinespace
        Baeuml, \newline et. al. \cite{baeuml2023security} & Cutoff \newline assumption & 12 & $10^{12} - 10^{15}$ & No & General  \newline(EAT) & $f=1-1.05$ \\ \addlinespace
        Pascual-García, \newline et. al. \cite{pascualgarcia2024} &  Cutoff \newline assumption & 12 & $10^{8} - 10^{10}$ & No & General  \newline(GEAT) & $f=1-1.05$\\ \addlinespace
        Primaatmaja, \newline et. al. \cite{primaatmaja2024} &  Dimension reduction & 12 & $10^{14} - 10^{16}$ & No & General  \newline(GEAT) & Shannon \newline limit \\ \addlinespace
        Navarro, \newline et. al. \cite{navarro2025} & Cutoff \newline assumption& 10 & $10^{6} - 10^9$ & No & General \newline (MEAT) & Shannon \newline limit  \\ 
        \midrule
        \textbf{This work} & Dimension reduction & 45 & $10^6-10^{10}$ & Yes & General (MEAT) & 0.95 \\ 
        \bottomrule
    \end{tabularx}
\end{table}

\section{Discussion} \label{Sec:Discussion}

We have derived a comprehensive finite-size security analysis for DM CVQKD under the MEAT framework \cite{arqand2025MEAT}. Our approach combines a MEAT-based security proof with several essential ingredients required for realistic protocol modeling. In particular, we consistently adapt the dimension reduction technique to enable the numerical estimation of secret key rates without a cutoff assumption, incorporate a trusted detector model that realistically accounts for inefficiencies and electronic noise in Bob’s setup, and apply postselection as a practical mechanism to mitigate the limitations imposed by the inherent vacuum noise in CV systems. Although our numerical analysis focuses on variable-length keys for the fully discretized QPSK protocol, the framework developed here applies to arbitrary discrete constellations of coherent states and can also be adapted to fixed-length key generation. For the latter, the key length is chosen in advance, and a statistical acceptance test ensures that Eq.~\eqref{eq: final key} holds for
every accepted announcement string.
% Although our numerical analysis focuses on the fully discretized QPSK protocol as a representative example, the framework developed here applies to arbitrary discrete constellations of coherent states. 

Our numerical evaluations explicitly demonstrate the advantages of this approach. 
We are able to extract positive key rates in the finite-size regime, specifically for a block size of $n = 10^{6}$, demonstrating that the MEAT-based analysis can overcome several limitations previously encountered in the security analysis of DM CVQKD protocols \cite{primaatmaja2024}. To contextualize these advancements, Table \ref{tab:qkd_comparison} provides a comparison of diverse references in the literature where the security analysis of DM CVQKD (particularly QPSK) is evaluated under different conditions. As illustrated, our work establishes a more comprehensive framework by replacing the cutoff assumption with the dimension reduction technique, while simultaneously incorporating detection imperfections.  
Crucially, these theoretical and practical improvements are achieved alongside an overall reduction in the required block size and a general, MEAT-based security proof that avoids imposing sequential limitations on the repetition rate. On the other hand, our framework makes use of trace norm-based continuity bounds \cite{Marwah_2022,Bluhm_2024}, which mathematically guarantee a strictly valid lower bound on the dimension reduction. Future work shall make use of recent advances in continuity bound based on fidelity to verify any possible source of improvement \cite{tan2026full}.

From a numerical perspective, low values of $N_c$ can be employed to minimize computational time while preserving unconditional security, with the only tradeoff being a more conservative key rate. Alternatively, a higher $N_c$ will yield a tighter approximation of the true infinite-dimensional state and therefore higher secret key rates. As this comes at a greater computational cost, we also derived a hyperplane approximation of the original conic optimization, which allows an increase of the effective dimension up to $N_c' = 45$ while keeping the computational cost tractable.

Overall, we note that our security proof places DM CVQKD at the same level as other protocols, such as decoy-state BB84 \cite{kamin25MEATsecurity}. In particular, as we have achieved a MEAT-based analysis for DM CVQKD that includes experimental imperfections, it allows a protocol-level description along the same lines as for discrete-variable protocols \cite{tupkary2026rigorouscompletesecurityproof,mizutani2025protocolleveldescriptionselfcontainedsecurity,tupkary2026meatsourcedetector}.

\textit{Note added--} Following an initial version of this work, Kamin et al. \cite{kamin2026unconditionalsecuritydiscretemodulatedcvqkd} provided a composable security proof for DM CVQKD with dimension reduction and identified technical issues in the security argument of this earlier version, which are addressed in the present manuscript. Our approach invokes a different dimension reduction construction and applies the finite-dimensional MEAT to a sequence of approximating states converging in trace norm to the physical state. This argument allows us to rigorously establish composable security without invoking an infinite-dimensional extension of the MEAT as in Ref. \cite{kamin2026unconditionalsecuritydiscretemodulatedcvqkd}, while achieving an exact formulation for the testing constraints.

\section*{Acknowledgments}
This project was supported by the Government of Spain (Severo Ochoa CEX2019-000910-S, FUNQIP and NextGeneration EU PRTR-C17.I1) and European Union (QSNP, 101114043). MN acknowledges funding from the European Union’s Horizon Europe research and innovation programme under the MSCA Grant Agreement No. 101081441. MA was supported by the Spanish Agencia Estatal de Investigación, Grant Nos.~RYC2023-044074-I and PID2024-161725OA-I00. CPG has received funding from the European Union’s Digital Europe Programme under the projects QUARTER (101091588) and QUARTERNEXT (101305103), and from the European Innovation Council's Horizon Europe EIC Accelerator Programme under the project MIQRO (101161539). AA acknowledges the ERC AdG CERQUTE, the AXA Chair in Quantum Information Science.

\printbibliography
\newpage
\appendix

\section{Privacy amplification via Rényi leftover hashing}\label{App:PrivacyAmp}

As a starting point for our finite-size analysis, we employ the leftover hash lemma \cite{tomamichel2015quantum,berta2016smooth} based on Rényi entropies \cite{Dupuis23RenyiHashLemma} considering that all registers are finite-dimensional. This lemma characterizes the distillable secret key length against quantum side information.

\begin{proposition}
    \emph{\cite[Theorem 9]{Dupuis23RenyiHashLemma}} \label{prop:RenyiHashLemma}
    Let $\sigma_{AE} \in \mc{D}(AE)$ be a cq-state and $\{\mc{R}^h_{A \rightarrow C}(\cdot), h \in \mc{H}\}$ a family of $\lambda-$randomizing hash functions on register $A$. Then, 
    \begin{equation}
        T \left(\rho^\mc{R}_{CFE}, \frac{1}{|C|} \mathds{1}_C \otimes \rho_{FE} \right) \leq 2^{\frac{2(1-\alpha)}{\alpha}}2^{\frac{\alpha-1}{\alpha}(\log|C|-H^\uparrow_\alpha(A|E)_\sigma + 2 \log \lambda)} ,\label{eq:renyihashlemma}
    \end{equation}
    where $\rho^\mc{R}_{CFE} = (\sum_h p(h) \mc{R}^h \otimes \pro{h}_F \otimes \mathds{1}_E)(\sigma_{AE})$, $\alpha \in (1,2)$ and $\{\ket{h}_F\}_{h \in \mc{H}}$ constitutes a basis in the register $F$ of hash functions.
\end{proposition}
To include this proposition in our analysis given a secret key of length $\ell(c_1^n)$, we condition on the observed announcement string, denoted by the event $\Omega_{c_1^n}:=\{C_1^n=c_1^n\}$, and successful verification $\Omega_{\mathrm{EV}}$. In addtion, we may apply the following notation substitutions\footnote{As noted in the main text, it is considered that the hash function also discards the postselection symbols.} 

\begin{align*}
    C &\rightarrow K_B, \\
    A &\rightarrow Z_1^n, \\
    E &\rightarrow E' L, \\
    \rho^{\mathcal{R}}_{CFE} &\rightarrow \rho_{K_B F E' L|\Omega_{c_1^n} \wedge \Omega_{\mathrm{EV}}}, \\ 
    \sigma_{AE} &\rightarrow \rho_{Z_1^n E' L|\Omega_{c_1^n} \wedge \Omega_{\mathrm{EV}}},\\
    \frac{1}{|C|} \mathds{1}_C\otimes \rho_{FE} &\rightarrow \bar{\tau}_{K_B F E' L|\Omega_{c_1^n} \wedge \Omega_{\mathrm{EV}}},
\end{align*}
where $L$ represents the string of classical bits leaked to the adversary by Alice and Bob during information reconciliation, $F$ stores the choice of the universal$_2$ hash function used during the postprocessing, and $E':= C_1^n C_{EV}\hat{E}$ contains Eve's global side information $\hat{E}$, the classical register $C_1^n$ with all public information associated with test rounds and the classical register $C_{EV}$ storing the public choice of the verification hash. Given the fact that we use ideal universal$_2$
% \cpg{Pequeña nota: los hashes que solemos utilizar a nivel industrial (Toeplitz) son ideales. Los hay que son algo más eficiente porque necesitan una pre-seed más pequeña (como el hash polinomial) pero que son delta-almost. En general la diferencia no se ve en papers teóricos mientras no entres en detalles sobre el QRNG} \m{a qué te refieres con delta almost?} \cpg{Eh hashes, los tienes perfectos (con colisión $1/l$ o $\delta-$almost (colisión $\delta/l$. El ideal es un $0-$almost}
hash functions, we have $\lambda=1$ \cite[Lemma 6]{Dupuis23RenyiHashLemma}. Now, recalling that Bob's final key is a binary register, whose length we will characterize as $\ell({c_1^n})$, we have that $|K_B| = 2^{\ell({c_1^n})}$. Thus, \eqref{eq:renyihashlemma} leads to

\begin{equation}
    T(\rho_{K_B F E'L|\Omega_{c_1^n} \wedge \Omega_{\mathrm{EV}}},\bar{\tau}_{K_B F E' L|\Omega_{c_1^n} \wedge \Omega_{\mathrm{EV}}}) \leq  2^{\frac{\alpha-1}{\alpha}[\ell-H^\uparrow_\alpha(Z_1^n| E' L)_{\rho_{|\Omega_{c_{1}^{n}} \wedge \Omega_{\mathrm{EV}}}}-2]}. \label{eq:TraceRenyi}
\end{equation}

Using \cite[Lemmas 5.14, 5.15]{tomamichel2015quantum}, we bound the effect of the leaked information as

\begin{equation} \label{eq:LeakRemoval}
    H^\uparrow_\alpha(Z_1^n| E' L)_{\rho_{|\Omega_{c_1^n} \wedge \Omega_{\mathrm{EV}}}}  \geq H^\uparrow_\alpha(Z_1^n| E' )_{\rho_{|\Omega_{c_1^n} \wedge \Omega_{\mathrm{EV}}}}  - \mathrm{leak}(c_1^n),
\end{equation}
where $\mathrm{leak}(c_1^n):=\log(|L|)$  bounds the total number of bits leaked during information reconciliation for the fixed observed announcement string $c_1^n$.

\section{Marginal-constrained entropy accumulation theorem} \label{app:MEAT}

In our analysis, we use the main result of \cite{arqand2025MEAT}, known as the marginal-constrained entropy accumulation theorem (MEAT). In particular, we consider an implementation where Alice and Bob do not disclose any information before applying their measurements to all quantum signals, which allows us to use a simplified version of the theorem. 

\begin{proposition}[Modified version of Corollary 4.2 and Theorem 4.1b from \cite{arqand2025MEAT}] \label{prop:MEAT} 
For each $j \in \{1,...,n\}$, take a state $\sigma^{j} \in \mc{D}(A_{j})$ and a CPTP map ${\mc{M}}_j: [AB]_{j} \to [ZCI]_j$, such that $C_j$ and $I_j$ are classical. Let $\rho_{[ZCI]^{n}_1 \hat{E}} = {\mc{M}}_n \circ ... \circ {\mc{M}}_1 (\bar{\omega}) $ for $\bar{\omega} \in \mc{D} ([AB]^{n}_1 \hat{E})$, verifying $\bar{\omega}_{A_1^{n}}= \sigma^{1}_{A_1}\otimes ... \otimes \sigma^{n}_{A_{n}}$.

Suppose furthermore that for all rounds we have a tradeoff function $f:\mc{C} \to \mathbb{R}$ and define the constant
\begin{equation}\label{eq:kappaInitial}
\begin{gathered}
  \kappa:= \inf_{\omega_{AB {E} }\in\mathcal{D}(\!AB{E}\!)} H_{\alpha}^{\uparrow,f} (Z | {C}I {E} )_{\mathcal{M}(\omega)}, \\
\text{s.t. }\Tr_{B}[\omega_{AB}]=\sigma_{A},
\end{gathered}
\end{equation}
where $E$ is a large-enough purifying register for any $[AB]_{j}$; and let the full, normalized tradeoff function
\begin{align}
\hat f(c_1^n)
:=
\sum_{j=1}^{n}\bigl(f(c_j)+\kappa\bigr)
=
n\bigl(f\cdot\mathrm{F}^\mathrm{obs}_{c_1^n}+\kappa\bigr),
\end{align}
where the observed frequency vector $\mathrm{F}^\mathrm{obs}_{c_1^n}$ is defined in Eq. \eqref{eq:frequency distr.}. Then, the corresponding $\hat{f}-$weighted Rényi entropy verifies
\begin{align}
    H^{\uparrow, \hat{f}}_\alpha (Z_1^n|[CI]^{n}_1 \hat{E})_\rho \geq 0,
\end{align}
for any $\alpha \in (1,\infty]$.
\end{proposition}

% My idea was to start with a finite-dimensional problem, for the optimization you can do $\kappa_m > \kappa$ where the first is the finite dimensional optimization problem and the second is the infinite dimensional. This is valid because you are enlarging the set which you can optimize, and since we are computing the infimum, that can only equals or lower the value of your optimization. We then apply dimension reduction and blabla. For the security, with the lemma here and Appendix A from \cite{tupkary2026rigorouscompletesecurityproof} you can state that secrecy for the finite-dimensional states is equivalent to secrecy for the infinite-dimensional states.

It is important to emphasize that this theorem is only defined for finite-dimensional registers. To accommodate the MEAT to our optical protocol, the following lemma constructs finite-dimensional approximations of the physical state:
% we ought to approximate the physical, infinite-dimensional state by finite-dimensional representations that preserve Alice's product marginal exactly while arbitrarily approximating the true state. 
% This is based on the fact that the underlying Hilbert space for phase space is separable, such that any infinite-dimensional state or operator is arbitrarily close to a finite version whose  dimension is large-enough. The following lemma constructs these
% approximations and establishes their trace-norm convergence.

\begin{lemma}[Finite-dimensional approximation preserving Alice's marginal]\label{lem:MEAT marginal approximation}
Let $\bar\omega\in\mc{D}([AB]_1^n\hat E)$ satisfy $\bar\omega_{A_1^n}=\sigma_A^{\otimes n}$, with $A$ finite-dimensional, and $B,\hat E$ separable. There exist a sequence of states $\bar\omega^{(m)}$ supported on the finite-dimensional space  $[AB^{(m)}]_1^n\hat E^{(m)}$, such that
\begin{equation}\label{eq:MEAT approximation properties}
    \lim_{m\rightarrow \infty}\norm{\bar\omega^{(m)}-\bar\omega}_1=0,
\end{equation}
which preserves the original marginal constraint, i.e., $\bar\omega^{(m)}_{A_1^n}=\sigma_A^{\otimes n}$ for every $m$.
\end{lemma}

\begin{proof}
We adapt the projection channels definition of \cite[Definition A.1]{tupkary2026rigorouscompletesecurityproof}, by mapping the discarded probability to a fixed state within the projected subspace, rather than to an additional orthogonal flag. For $X\in \{B_1,..., B_n,\hat E\}$, choose the nested finite-rank projectors $P_X^{(m)}$ which converge strongly to $\id_X$, and a unit vector $\ket{0_X}$ in every retained space. On every register $B_j$, take the same Fock projector $P_B^{(m)}=\sum_{n=0}^{m}\pro{n}_B$. Set
\begin{align}
    \mc{C}_X^{(m)} (\rho) 
    &=P_X^{(m)}\rho P_X^{(m)}+\Tr[(\id_X-P_X^{(m)})\rho]\pro{0_X},\label{eq:MEAT compression channel}\\
    \bar\omega^{(m)}
    &=\left(\operatorname{id}_{A_1^n}\otimes\bigotimes_{j=1}^n\mc{C}_{B_j}^{(m)}\otimes\mc{C}_{\hat E}^{(m)}\right)(\bar\omega).\label{eq:MEAT compressed block}
\end{align}
These maps are CPTP and act trivially on $A_1^n$, so Alice's marginal is unchanged. Defining $Q_m=\id_{A_1^n}\otimes\bigotimes_j P_{B_j}^{(m)}\otimes P_{\hat E}^{(m)}$, we expand the product of channels, where we obtain $\bar\omega^{(m)}=Q_m\bar\omega Q_m+R_m$, with $R_m\succeq0$ and $\Tr[R_m]=1-\Tr[Q_m\bar\omega]$. Hence, applying the triangle inequality, we have 
\begin{equation}\label{eq:MEAT block approximation error}
    \norm{\bar\omega^{(m)}-\bar\omega}_1 = \norm{Q_m \bar \omega Q_m - \bar\omega + R_m}_1
    \leq\norm{Q_m\bar\omega Q_m-\bar\omega}_1+1-\Tr[Q_m\bar\omega]
\end{equation}
which tends to zero as $m\to\infty$ by trace-norm convergence of finite projections \cite[Section VI, Theorem 4.3]{gohberg1990classes}.
\end{proof}

For a finite $m$,  we apply Proposition \ref{prop:MEAT} to the CPTP restriction $\mc{M}^{(m)}:AB^{(m)}\rightarrow ZC$ of the complete channel $\mc{M}$. 
% The constant $\kappa$ is understood as the value in Eq. \eqref{eq:kappaInitial} for this restricted channel, with its dependence on $m$ left implicit. 
For later reference, we also define
\begin{equation}\label{eq:MEAT supported marginal set}
    \Sigma_\Omega^{(m)}
    =\{\omega\in\Sigma_\Omega:(\id_A\otimes P_B^{(m)})\omega(\id_A\otimes P_B^{(m)})=\omega\}.
\end{equation}
where $\Sigma_\Omega
=
\left\{
\omega_{AB}\in\mathcal D(AB):
\operatorname{Tr}_B[\omega_{AB}]=\sigma_A
\right\}$ is the set of infinite-dimensional states. 
% % constraint $\omega_{AB}\in\Sigma_\Omega^{(m)}$. 
% Every $\omega_{AB}$ admits a
% purification with a purifying register of dimension at most
% $\dim(A)\dim(B^{(m)})$. Any other extension of the same state is obtained by a channel on the purifying register, so data processing shows that using a purification cannot increase the objective. Its value is independent of the chosen purification by isometric invariance. Thus the finite-dimensional problem has the same objective and a smaller feasible set, so that 

% for every $m$. Here, $\kappa$ optimizes over all states on the original Hilbert spaces, without a finite-dimensional support restriction.
\section{Dimension reduction}\label{app:DimRed}

\subsection{Bounds on the sandwiched Rényi entropy}

In order to achieve an expression that is well-defined for later numerical calculations, we use the following continuity bounds for sandwiched Rényi entropies.

\begin{proposition} \emph{\cite{Marwah_2022,Bluhm_2024}}  \label{prop:renyi continuity bound}
    Let $\alpha \in [\frac{1}{2},1)$, $\epsilon \in [0,1]$ and $\rho,\sigma \in \mc{D}(AB)$ where $A \in \mathcal{A}$ is a classical register. If $T(\rho_{AB},\sigma_{AB}) \leq \epsilon$, then
    \begin{equation} \label{eq:ContBoundsLow}
        |H^\uparrow_\alpha(A|B)_\sigma - H^\uparrow_\alpha(A|B)_\rho | \leq \log(1+\epsilon) + \frac{1}{1-\alpha} \log \left( 1 + \epsilon^\alpha |\mathcal{A}|^{1-\alpha} - \frac{\epsilon}{(1+\epsilon)^{1-\alpha}} \right).
    \end{equation}
On the other hand, for $\alpha >1$
    \begin{equation} \label{eq:ContBoundsHigh}
        |H^\uparrow_\alpha(A|B)_\sigma - H^\uparrow_\alpha(A|B)_\rho | \leq \zeta_\alpha(\epsilon),
    \end{equation}
    with
    \begin{align} \label{eq:zeta function}
        \zeta_\alpha(\epsilon) = \min \begin{cases}
            \log(1+\epsilon) + \frac{1}{\alpha-1} \log \left( 1 + \epsilon |\mc{A}|^{\alpha-1} - \frac{\epsilon^\alpha}{(1+\epsilon)^{\alpha-1}} \right), \\
            \frac{\alpha}{\alpha-1} \log \left( 1 + \epsilon |\mc{A}|^{(\alpha -1)/\alpha}\right), \\
            \log(1+\epsilon) + \frac{\alpha}{\alpha - 1} \log \left( 1 + \epsilon |\mc{A}|^{(\alpha - 1)/\alpha} - \frac{\epsilon^{2 - \frac{1}{\alpha}}}{(1+\epsilon)^{\frac{\alpha-1}{\alpha}}}\right).
        \end{cases}
    \end{align} \label{Prop:ZetaBounds}
\end{proposition}
\begin{remark}
Compared to \cite{Bluhm_2024}, we observe that this bound takes $|\mc{A}|^2 \to |\mc{A}|$ since we assume that register $A$ is classical \cite[Section XII]{kamin25MEATsecurity}. 
\end{remark}

Let us set a finite, arbitrarily large dimension $m$ in the approximation of Appendix \ref{app:MEAT} and consider a single-round state $\omega_{AB}\in\Sigma_\Omega^{(m)}$ with a purification $\omega_{ABE}$. We omit the index $m$ on this state to keep the notation simple. Next, we introduce

\begin{equation}\label{eq:finite dim state}
    \tilde \omega_{ABE} = \Pi \omega_{ABE} \Pi,
\end{equation}
where we define the projector $\Pi = (\id_A \otimes \Pi_B \otimes \id_E)$ with $\Pi_B = \sum_{n=0}^{N_c} \pro{n}_B$ for $N_c \in \mathbb{N}$. Note that $m>N_c$.
Since \(\tilde\omega_{ABE}\) is pure, it holds
\begin{equation}\label{eq:projected purification rank}
    \operatorname{rank}\tilde\omega_E
    =\operatorname{rank}\tilde\omega_{AB}
    \leq\dim(A)(N_c+1).
\end{equation}
 This argument is also valid when the original purification has an infinite-dimensional register $E$. We also define $\Pi'= \id-\Pi$ and the normalized pinched state
\begin{equation} 
    \hat\omega_{ABE} = \Pi \omega_{ABE}\Pi + \Pi' \omega_{ABE}\Pi'.
\end{equation}
Here, $\hat\omega$ remains finite-dimensional for each $m$, but its support retains components greater than the cutoff $N_c$. Consequently, it cannot be represented only within the truncated space defined by $\Pi_B$. To express its contributions, we define $\hat\Pi^{c} \in \{\id_A\otimes R^\top, \proj{0}\otimes \tilde{R}_B^0,...,\proj{3}\otimes\tilde{R}^8_B\}$ with $c \in \mc{C}$, together with 
\begin{align}
    s(c) &:= \Tr[\hat\Pi^c \Pi' \omega \Pi'], \label{eq: s(c)} \\
    t(c) &:= \mathrm{Re}\{\Tr[\hat\Pi^c \Pi \omega \Pi']\}. \label{eq:t(c)}
\end{align}
Thus, we can express the one-round channel in Eq. \eqref{eq:CompleteMap} as
\begin{equation}\label{eq:OneRoundCh}
    \mc{M}(\omega)_c =  \bar{p}(c) \Tr[\hat\Pi^c \omega] = \bar{p}(c)(\Tr[\hat\Pi^c \tilde\omega] +  s(c) + 2 t(c)). 
\end{equation}
Using the fact that $R^\top$ is diagonal in the Fock basis, we have that $[R^\top_B, \Pi_B]=0$. Then, $\Tr[R^\top \omega]=\Tr[R^\top \hat\omega]= \Tr[R^\top \tilde\omega] + s(\perp)$ where

\begin{align} \label{eq:s(bot)}
    s(\bot) = \sum_{x=0}^3 \sum_{z=4}^7 s(x,z).
\end{align}
We now apply Proposition \ref{prop:renyi continuity bound} to $\widetilde{\mc{M}}(\omega)_\perp$ and $\widetilde{\mc{M}}(\hat\omega)_\perp$.
These states have the same normalization given by 
\begin{align}\label{eq:KeyMapDecomp}
    \Tr[\mc{M}(\omega)_\perp] = \pkey \left(\Tr[R^\top \tilde\omega] + s(\perp)\right),
\end{align} and their difference
depends only on
$\omega-\hat\omega=\Pi\omega\Pi'+\Pi'\omega\Pi$. Let us define $\tilde{t}(x,z):= \mathrm{Re}\{\Tr_{AB}[(\hat\Pi^{(x,z)}) \otimes \id_E) \Pi \omega \Pi']\}$. Then,

\begin{align} \label{eq: final penalization argument}
    T(\widetilde{\mc{M}}(\omega_{ABE})_\perp,\widetilde{\mc{M}}(\hat\omega_{ABE})_\perp)  &= \frac{\sum_{z=4}^7\norm{\sum_{x=0}^3 \tilde{t}(x,z)}_1}{\Tr[R^\top \omega]} 
    \nonumber \\
    &\leq \frac{\sum_{z=4}^7\sum_{x=0}^3\norm{ \tilde{t}(x,z)}_1}{\Tr[R^\top \omega]}
    \nonumber \\
    &\leq\frac{\sum_{z=4}^7\sum_{x=0}^3\sqrt{ \Tr[\hat\Pi^{(x,z)}\tilde{\omega}] s(x,z)}}{\Tr[R^\top \tilde\omega] + s(\perp)} 
    := k,
\end{align}
where the second line uses the triangle inequality and the third line uses the inequality in \cite[Eq. 1.183]{Watrous2018} together with the definition of the trace norm considering $\omega_{ABE}$ is pure. 
Applying the bound \eqref{eq:ContBoundsHigh} we find

\begin{align}\label{eq:ContinuityFinal}
    H^\uparrow_{\alpha} (Z|E)_{\widetilde{\mc{M}}(\omega)_\perp} \geq  H^\uparrow_{\alpha}(Z|E)_{\widetilde{\mc{M}}(\hat\omega)_\perp} - \zeta_\alpha (k).
\end{align}
Regarding $\widetilde{\mc{M}}(\hat\omega)_\perp$, we note its special structure due to the pinching acting on Bob's
\begin{align}
    \widetilde{\mc{M}}(\hat\omega)_\perp &= \widetilde{\mc{M}}(\Pi\omega \Pi + \Pi' \omega \Pi')_\perp \nonumber\\
    % &= \sum_{z=0}^3 \Tr_{AB}[\mathds{1}_{AE} \otimes R^z_B (\Pi \omega_{ABE} \Pi + \Pi' \omega_{ABE} \Pi') \pro{z}_Z \\
    &= \sum_{z \in \hat{\mc{Z}}} \pro{z}_Z \{ \Tr_{AB}[\mathds{1}_{AE} \otimes \Pi R^z_B \Pi ( \Pi \omega_{ABE}) \Pi] + \Tr_{AB}[(\mathds{1}_{AE} \otimes \Pi' R^z_B \Pi' ) \Pi' \omega_{ABE} \Pi']) \} \nonumber \\
    &=:  \sum_{z \in \hat{\mc{Z}}} \pro{z}_Z \Tr_{AB}[(K^z)^\dagger K^z \hat{\omega} ] 
\end{align}
where in the last line we denoted the POVM elements in an isometric fashion
\begin{align}
    K^z = \mathds{1}_{AE} \otimes \begin{pmatrix}
        \sqrt{\Pi R^z \Pi} & 0 \\
        0 & \sqrt{\Pi' R^z \Pi'}
    \end{pmatrix} =: \mathds{1}_{AE} \otimes \begin{pmatrix}
        K^z_\Pi & 0 \\
        0 & K_{\Pi'}^z
    \end{pmatrix},
\end{align}
which together define the coherent key map $\g$ provided by the superoperator
\begin{align}
    G = \sum_{z \in \hat{\mc{Z}}} \ket{z}_R \otimes \mathds{1}_A \otimes {K^z_B}.
\end{align}
We also define the dephasing map on $R$ by
\begin{align}
    Z_r = \pro{r}_R \otimes \id_{AB}.
\end{align}
Let $\ket{\hat{\omega}}_{ABEJ}$ be a purification of the pinched state $\hat{\omega}_{ABE}$ and define $\widetilde{\mathcal{N}}:= (\Tr_{AB} \circ \zg) \otimes \id_{EJ}$, such that $\Tr_J[{\widetilde{\mathcal{N}}(\hat{\omega}_{ABEJ})}]=\widetilde{\mc{M}}(\hat{\omega}_{ABE})_\perp$. Hence, using the data processing inequality and \cite[Theorem 2]{Chung_2025}, we have
\begin{align}
    H^\uparrow_\alpha(Z|E)_{\widetilde{\mc{M}}(\hat{\omega}_{ABE})_\perp} \geq H^\uparrow_\alpha(Z|EJ)_{\widetilde{\mathcal{N}}(\hat{\omega}_{ABEJ})_\perp} \geq D_\mu( \g(\hat\omega_{AB}) || \zg(\hat\omega_{AB})),
\end{align}
with $\mu = 1/\alpha$. 
% Here, we use the decomposition
% \begin{align}
%     \rho_{ZRAB} &= V \rho_{RAB} V^\dagger =V \g( \hat{\omega})V^\dagger,
% \end{align}
% to denote the state shared by Alice, Bob, and Eve after applying the key map, which generates an intermediate register $R$, and the final key extraction via the isometry $V$ by acting on said register. In particular,
% \begin{align}\label{eq:DephIsometry}
%     V = \sum_{z \in \hat{\mc{Z}}} \ket{z}_Z \otimes Z^z_R,
% \end{align}
% with $\{Z^z\}_{z \in \hat{\mc{Z}}}$ representing a projective measurement. Provided these considerations, and following \cite[Appendix A]{navarro2025}, the entropy above leads to
% \begin{align}
%     H^\uparrow_\alpha(Z|E)_{\widetilde{\mc{M}}(\hat{\omega})_{\bot}} &\geq D_\mu (\rho_{RAB} || \sum_{z} Z^z_R \rho_{RAB} Z^z_R) = D_\mu (\g(\hat{\omega}) || \zg(\hat{\omega})).
% \end{align}
% where we observe the dephasing $\z(\cdot) = \sum_{r =0}^3 Z_r (\cdot) Z^\dagger_r$, which results from the isometry \eqref{eq:DephIsometry} and is described by the superoperators
% \begin{align}
%     Z_r = \pro{r}_R \otimes \id_{AB}.
% \end{align}
Now we note that Bob's pinching and the block-diagonal nature of $\g$ induce a division
\begin{align}
    \g(\hat{\omega}) = \g_\Pi (\Pi \omega \Pi) + \g_{\Pi'} (\Pi' \omega \Pi'),
\end{align}
which we can extend to the $\z$ map by linearity. These two components are orthogonal at Bob's space, and the pinching does not act on register $B$, so we may split the total function according to said orthogonality
\begin{align}
    &D_\mu (\g(\hat{\omega}), \zg(\hat{\omega})) = \frac{1}{\mu-1}\log\left[ \frac{\Psi(\g(\hat{\omega}), \zg(\hat{\omega}))}{\Tr[\g(\hat{\omega})]} \right]  \nonumber \\
    &\quad = \frac{\alpha}{1-\alpha}\log\left[ \frac{\Psi_\mu(\g_\Pi(\Pi \omega \Pi) , \zg_{\Pi}(\Pi \omega \Pi))) +  \Psi_\mu (\g_{\Pi'}(\Pi' \omega \Pi'), \zg_{\Pi'}(\Pi' \omega \Pi'))}{\Tr[\g(\hat{\omega})]} \right].
\end{align}
In the case of the infinite-dimensional component, we simply apply Hölder's inequality together with the trace-preserving nature of the pinching to find 
\begin{align} \label{eq:HoelderInfPsi}
    \Psi_\mu (\g_{\Pi'}(\Pi' \omega \Pi') || \zg_{\Pi'}(\Pi' \omega \Pi')) &\leq \Tr[G_{\Pi'}(\Pi' \omega \Pi' )] = s(\bot).
\end{align}
As a result, the remaining component of the Rényi relative entropy cone is finite-dimensional and amenable for numerical analysis.

\subsection{Bounds on the marginal}
 We adapt the original marginal condition $\Tr_B[\omega_{AB}] = \sigma_A$, which also requires an infinite-dimensional representation. Provided that the state $\ket{\omega}_{ABE}$ is pure, it holds \cite[Appendix A]{primaatmaja2024}
\begin{align} \label{eq: marginal decomposition}
    \sigma_A = \Tr_{BE}[\tilde\omega_{ABE}] + \Tr_{BE}[\Pi'\omega_{ABE}\Pi'], 
\end{align}
where the second term of the last equality is positive semidefinite with trace $1-\Tr[\tilde{\omega}]$, but otherwise unknown. Hence, every admissible state $\tilde\omega_{ABE}$ satisfies
\begin{equation}
    \sigma_A -\Tr_{BE}[\tilde\omega_{ABE}] \succeq 0,
\end{equation}
which is equivalent to $\norm{\sigma_A -\Tr_{BE}[\tilde\omega_{ABE}] }_1 = 1-\Tr[\tilde{\omega}]$. 

\subsection{Bounds on the region operators}\label{Subsec:StructuralConstraints}

Let us apply an effective dimension reduction for the POVM elements. From \eqref{eq:OneRoundCh} we can reduce their action into a compact support, but leading to a suboptimal protocol performance since the infinite-dimensional components are fundamentally unbounded. We first consider $s(c)$ (see Eq. \eqref{eq: s(c)}), which describes the contribution above the cutoff. We start by imposing
\begin{align}
    s(c) \geq 0, \quad \forall c \in \mc{C}.
\end{align}
which stems from the positivity of the inner product for any two positive semidefinite operators. 
Along the lines of \cite{LinTrusted2020,Kanitschar2023}, we can fin the following expression for sums of the region operators
\begin{align}
   \sum_{z=0}^3\tilde R_B^z
    &=
    \sum_{n=0}^{\infty}
    \sum_{j=0}^{n}
    \begin{pmatrix}
        n \\ j
    \end{pmatrix}\frac{C_n}{a^{j+1} b^j j!}
     \int_0^{a\Delta_s^2} u^j e^{-u} du
    \pro{n}, 
    \label{eq:analytical sumR03}\\
    \sum_{z=4}^7 \tilde{R}^z_B = R^\top_B &= \sum_{n=0}^\infty   \sum_{j=0}^n \begin{pmatrix}
        n \\ j
    \end{pmatrix}\frac{C_n}{a^{j+1} b^j j!} \int_{a\Delta_s^2}^{a\Delta^2} u^j e^{-u} du \pro{n}, 
    \label{eq:analytical sumR47}\\
    \tilde{R}^8_B &= \sum_{n=0}^\infty  \sum_{j=0}^n \begin{pmatrix}
        n \\ j
    \end{pmatrix} \frac{C_n}{a^{j+1} b^j j!} \int_{a\Delta^2}^{\infty} u^j e^{-u} du\pro{n}.
    \label{eq:analytical R8} 
\end{align}
Here, we used the change of variable $u=ar^2$, and we conveniently set
\begin{subequations}\label{eq:NoiseCoeffs}
\begin{align}
    \bar{n} &= \frac{1-\eta_d + \nu}{\eta_d}, \\
    a &= \frac{1}{\eta_d (1 + \bar{n})}, \\
    b &= \eta_d \bar{n} (1 + \bar{n}), \\
    C_n &= \frac{\bar{n}^n }{ \eta_d (1+\bar{n})^{n+1}}.
\end{align}
\end{subequations}
These operators are related by $\sum_{z=0}^3 \tilde{R}^z_B = \id_B - R^\top_B - \tilde{R}^8_B$ with $ R^\top = \sum_{z=4}^7 \tilde{R}^z_B$. Consequently, using Eq. \eqref{eq: marginal decomposition} we can express
\begin{align}\label{eq:PartialTraceSum}
    \sum_{z=0}^8 s(x,z)
    =\mathrm{Tr}\big[(|x\rangle\langle x|_A\otimes \mathds{1}_B)\,\Pi'\omega\Pi'\big]=\langle x|\,\mathrm{Tr}_B[\Pi'\omega\Pi']\,|x\rangle
    =\langle x|\sigma_A-\mathrm{Tr}_B[\tilde\omega]|x\rangle .
\end{align}
Thus, Alice's marginal fixes the total contribution above the cutoff for each of her outcomes. It does not, however, specify how this contribution is distributed among the inner, middle, and outer regions. Since all these sums arise from the same positive operator $\bra{x}\Pi'\omega\Pi'\ket{x}$, their distribution is further restricted by the region operators. 

We therefore need a bound on the inner contribution relative to the middle, of the form
\begin{equation}\label{eq:sInnerMiddleTarget}
    \sum_{z=0}^{3}s(x,z)
    \leq r\sum_{z=4}^{7}s(x,z),
    \qquad x\in\{0,\ldots,3\},
\end{equation}
for some unknown value $r$. To preserve positive semidefiniteness, a sufficient condition is  
\begin{equation}\label{eq:InnerMiddleOperatorBound}
    (\id_B-\Pi_B)
    \Big(\sum_{z=0}^{3}\tilde R_B^z \Big)
    (\id_B-\Pi_B)
    \preceq
    r(\id_B-\Pi_B)R_B^\top(\id_B-\Pi_B).
\end{equation}
Indeed, taking its trace against $\bra{x}\Pi'\omega\Pi'\ket{x}$ we recover \eqref{eq:sInnerMiddleTarget}. Moreover, \eqref{eq:InnerMiddleOperatorBound} is equivalent to 
\begin{equation}
    \sum_{z=0}^{3}\bra{n}\tilde R_B^z\ket{n}
    \leq r\bra{n}R_B^\top\ket{n},
\end{equation}
for every $n>N_c$. Since the diagonal entries of $R^\top_B$ are strictly positive, we can define 
\begin{equation}
    r_\text{in}(n) := \frac{ \sum_{z=0}^3 \bra{n}\tilde{R}^z_B\ket{n}}{\bra{n} R^\top_B\ket{n}}.
\end{equation}
Then, we can find $r = \sup_{n>N_c} r_\text{in}(n)$. To evaluate this supremum, we show that $r_\text{in}(n)$ is nonincreasing. It is sufficient to take the numerator when evaluating $r_\text{in}(n) - r_\text{in}(n+1)$ and prove that is positive, such that
\begin{align} \label{eq: difference inner ratios}
    \Big(&\sum_{z=0}^{3} \bra{n}\tilde R_B^z\ket{n} \Big) \bra{n+1}R_B^\top\ket{n+1}
    - \Big(\sum_{z=0}^{3} \bra{n+1} \tilde R_B^z\ket{n+1} \Big)\bra{n}R_B^\top\ket{n} \nonumber \\
    &= \sum_{0\leq k < j}^{n+1}\frac{C_n C_{n+1}}{a^2 (ab)^{j+k}j!k!} \left[ \binom{n+1}{j} \binom{n}{k}- \binom{n+1}{k}\binom{n}{j}\right] \int_0^{a\Delta_s^2} \int_{a\Delta_s^2}^{a\Delta^2} e^{-u-v}(u^k v^j - u^j v^k) dv du.
\end{align}
where we replaced the definitions of the region operators' sums. Note that we can rewrite
\begin{align}
    \binom{n+1}{j} \binom{n}{k}- \binom{n+1}{k}\binom{n}{j} = \frac{j-k}{n+1}\binom{n+1}{j}\binom{n+1}{k}
\end{align}
which is positive given that $k<j$. Furthermore, the integration domain satisfy $0 \leq u \leq v$, and hence
\begin{equation}
    u^k v^j - u^j v^k = u^k v^k(v^{j-k} - u^{j-k}) \geq 0.
\end{equation}
Thus, the difference in \eqref{eq: difference inner ratios} is nonnegative, which proves that $r_\text{in}(n) \geq r_\text{in}(n+1)$. It follows that 
\begin{equation}
    r=\sup_{n>N_c}r_{\mathrm{in}}(n)
    =r_{\mathrm{in}}(N_c+1),
\end{equation}
giving us the constraint
\begin{align}\label{eq:sInnerMiddle}
    &\bra{N_c+1}R_B^\top\ket{N_c+1}
    \sum_{z=0}^{3}s(x,z)
\leq
    \left(
        \sum_{z=0}^{3}
        \bra{N_c+1}\tilde R_B^z\ket{N_c+1}
    \right)
    \sum_{z=4}^{7}s(x,z),
\end{align}
for every $x\in\{0,\ldots,3\}$. 

We next bound the outer contribution from below relative to the middle contribution, through
\begin{align}
    \bra{n}\tilde{R}^8_B \ket{n} \geq r'  \bra{n}\tilde{R}^\top_B \ket{n},
\end{align}
where $r' := \inf_{n>N_c} r_\text{out}(n)$, with
\begin{align}
     r_{\mathrm{out}}(n)
    :=
    \frac{
        \bra{n}\tilde R_B^8\ket{n}
    }{
        \bra{n}R_B^\top\ket{n}
    }.
\end{align}
Following a similar procedure as before, we can also show that $r_\text{out}(n+1) \geq  r_\text{out}(n)$. This leads us to the constraint
\begin{align}
    \bra{N_c+1}\tilde R_B^8\ket{N_c+1}\sum_{z=4}^7s(x,z) \leq
 \bra{N_c+1}R_B^\top\ket{N_c+1}\,s(x,8),\quad  x \in \{0,...,3\}.
\end{align}
We conclude by using \cite[Lemma 5]{primaatmaja2024}
\begin{align}
    |t(c)| \leq \sqrt{\Tr[\hat{\Pi}^c \tilde{\omega}] s(c) }, \quad \forall c \in \mc{C},
\end{align}
which bounds the value for the off-diagonal contributions of the operators. Similarly, one can observe that 
\begin{align}
    t(x,8)=\sum_{z=0}^3 t(x,z) = \sum_{z=4}^7 t(x,z) = 0 \quad x \in \{0,...,3\},
\end{align}
which stems from the fact that the corresponding POVM elements (or the sum thereof) constitute diagonal operators, such that the off-diagonal components vanish by definition.

\section{Facial reduction} \label{app:Facial}
The key map for the truncation is

\begin{align}\label{eq:Gmap}
    G_\Pi &= \sum_{r =0}^3 \ket{r}_R \otimes \id_A \otimes \sqrt{\Pi R^r_B \Pi}.
\end{align}
As shown in \cite{navarro2025}, the map $\g_\Pi$ is not strictly positive, and thus necessarily requires a facial reduction to obtain a well-defined map to perform numerical optimization (which ensures strict feasibility \cite{Nesterov2018}). This new map $\gmap$, obtained after performing said procedure, can be derived from the support of $\g$, and is provided by the Kraus operator

\begin{align}
\hat{G} = \id_A \otimes \sqrt{\Pi R^\top_B \Pi},
\end{align}
where we have that $R^\top_B= \sum_{r=0}^3 R^r_B$ is diagonal in the Fock basis. Both maps are related via an isometry $W$ that satisfies $\g (\cdot) =W \gmap (\cdot) W^\dagger$ and $W^\dagger W = \id_{AB}$, where
\begin{align} \label{eq:IsometryW}
W = \sum_{r =0}^3 \ket{r} \otimes \id_A \otimes \left[\sqrt{\Pi R^r_B \Pi} \left(\sqrt{\Pi R^\top_B \Pi}\right)^{-1} \right].
\end{align}
On the other hand, the map $\z \circ \g_\Pi$ is strictly positive and thus coincides with its facial reduction $\zgmap$ \cite{lorente2024}. Provided these changes, we reformulate the function $\Psi_\mu$ corresponding to the truncated representation as
\begin{align} \label{eq:PsiNotation}
    \hat{\Psi}_\mu(\tilde{\omega}) := \Psi_\mu(\g_\Pi(\Pi \omega \Pi) , \zg_{\Pi}(\Pi \omega \Pi))),
\end{align}
to ease the notation.

\section{Fixed-length secret key rates} \label{app:FixedLength}

Starting from the finite-dimensional program that defines the fixed-length secret key rate \cite{arqand2025MEAT,navarro2025} which eventually becomes necessary to define the tradeoff function (see Appendix \ref{App:VarLengthKeys})
\begin{align} \label{eq:original optimization}
    h_\alpha := \inf_{{q}\in S_{\Omega}} \inf_{\omega \in \Sigma_\Omega^{(m)}} \frac{\alpha}{\alpha-1} D_\mathrm{KL}({q}\|\mc{M}(\omega_{AB})_{C}) + q(\perp ) H^\uparrow_\alpha (Z|E)_{\widetilde{\mc{M}} (\omega)_{|\perp}},
\end{align}
where we have the set of accepted, observed frequencies $S_\Omega$, and that the KL divergence can be decomposed as
\begin{align}
    D_{KL}({q}\|\mc{M}(\omega_{AB})_{C}) =  \sum_{c \in \tilde{\mc{C}}} q(c) \log \left(\frac{q(c)}{\mc{M}(\omega_{AB})_{c}} \right) + q(\perp) \log \left(\frac{q(\perp)}{\mc{M}(\omega_{AB})_{\perp}} \right),
\end{align}
from \eqref{eq:OneRoundCh} we explicitly decompose the values for genuine test rounds as
\begin{align}
    \sum_{c \in \tilde{\mc{C}}} q(c) \log \left(\frac{q(c)}{\mc{M}(\omega_{AB})_{c}} \right) = \sum_{c \in \tilde{\mc{C}}} q(c) \log \left( \frac{q(c)}{\bar{p}(c)(\Tr[\hat\Pi^c \tilde\omega] +  s(c) + 2 t(c))} \right).
\end{align}
On the other hand, we gather the $ \perp$-dependent terms of the objective function to achieve
\begin{align} \label{eq:perp part of the D_KL}
   h_\alpha^\perp := q(\perp) \left[ \frac{\alpha}{\alpha-1} \log \left(\frac{q(\perp)}{\mc{M}(\omega_{AB})_{\perp}} \right) + H^\uparrow_\alpha (Z|E)_{\widetilde{\mc{M}}(\omega)_{|\perp}} \right].
\end{align}
We note that 
\begin{align}
    q(\perp)  H^\uparrow_\alpha (Z|E)_{\widetilde{\mc{M}} (\omega)_{|\perp}} = - \frac{\alpha}{\alpha-1} q(\perp) \log \left (2^{-\frac{\alpha-1}{\alpha}H^\uparrow_\alpha (Z|E)_{\widetilde{\mc{M}} (\omega)_{|\perp}}} \right).
\end{align}
Replacing this in Eq. \eqref{eq:perp part of the D_KL} and considering \eqref{eq:KeyMapDecomp}, we arrive at
\begin{align} \label{eq: perp term obj function simplified}
    h_\alpha^\perp =  \frac{\alpha}{\alpha-1} q(\perp) \log \left( \frac{q(\perp)}{2^{-\frac{\alpha-1}{\alpha}H^\uparrow_\alpha (Z|E)_{\widetilde{\mc{M}} (\omega)_{|\perp}}} \pkey   \Tr[\g(\omega)]} \right).
\end{align}
Next, we use the continuity bound from \eqref{eq:ContinuityFinal} to project $\omega$ into its pinched version $\hat{\omega}$. The entropy corresponding to the truncated state can be bounded using the derivations from Appendix \ref{app:DimRed} to find
\begin{align}
    \Tr[\g(\omega)]\,2^{-\frac{\alpha-1}{\alpha}H^\uparrow_\alpha (Z|E)_{\widetilde{\mc{M}} (\hat\omega)_{|\perp}}}
    &\leq \Psi_\mu\big(\g(\hat\omega),\zg(\hat\omega)\big) \nonumber\\
    &= \Psi_\mu\big(\g_\Pi(\tilde\omega),\zg_\Pi(\tilde\omega)\big)+\Psi_\mu\big(\g_{\Pi'}(\Pi'\omega\Pi'),\zg_{\Pi'}(\Pi'\omega\Pi')\big) \nonumber\\
    &\leq \hat\Psi_\mu(\tilde\omega_{AB}) + s(\perp), \label{eq:infinityBound}
\end{align}
where in the last line we use the representation for $\Psi_\mu$ after performing the facial reduction (see Appendix \ref{app:Facial}), and \eqref{eq:HoelderInfPsi}. Taking the first branch of \eqref{eq:zeta function} as the minimum, we can express
\begin{align}
    2^{\frac{\alpha-1}{\alpha}\zeta_\alpha^{(1)}(k)} 
    = (1 + k)^\frac{\alpha-1}{\alpha} 
    \left(1 + k |\hat{\mc{Z}}|^{\alpha-1}- \frac{k^\alpha}{(1+k)^{\alpha-1}}\right)^\frac{1}{\alpha}.
\end{align}
\begin{remark}
    We use this branch as extensive numerical research finds that it is actually the minimum for any relevant range of values in $\alpha$ and $k$. However, any of the other branches actually gives a valid lower bound since taking a suboptimal branch can only induce an extra penalty in the final key rate.
\end{remark}
Multiplying by $\Tr[\g(\omega_{AB})]$, let 
\begin{align}\label{eq: simplified penalization}
    \tilde{\zeta}_\alpha(k) := \Tr[\g(\omega_{AB})]2^{\frac{\alpha-1}{\alpha}\zeta_\alpha^{(1)}(k)}.
\end{align}
Hence, we find a lower bound for Eq. \eqref{eq: perp term obj function simplified}, such that
\begin{align}\label{eq: perp term obj function second}
    h_\alpha^\perp \geq \bar h_\alpha^\perp  
    = 
    \frac{\alpha}{\alpha-1} q(\perp) \log 
    \left( 
    \frac{q(\perp) \Tr[\g(\omega_{AB})]}{\pkey \tilde{\zeta}_\alpha(k) (\hat\Psi_\mu(\tilde\omega_{AB}) + s(\perp))
    }
    \right).
\end{align}
Using Hölder's inequality, we can show that $\hat\Psi_\mu(\tilde\omega) \leq \Tr[R^\top \tilde\omega] $, and since $\zeta_\alpha \geq0$, we have $\tilde{\zeta}_\alpha(k) \geq \Tr[\g(\omega_{AB})]$. Therefore, we can find the expansion
\begin{align}
    \frac{\tilde{\zeta}_\alpha(k)}{\Tr[\g(\omega_{AB})]} (\hat\Psi_\mu(\tilde\omega_{AB}) + s(\perp))
   &=\hat\Psi_\mu(\tilde\omega_{AB}) + s(\perp) + \Big(\frac{\tilde{\zeta}_\alpha(k)}{\Tr[\g(\omega_{AB})]}-1\Big)\big(\hat\Psi_\mu(\tilde\omega_{AB}) + s(\perp)\big) \nonumber\\
    &\leq 
    \hat\Psi_\mu(\tilde\omega_{AB})+\tilde{\zeta}_\alpha(k) - \Tr[R^\top \tilde\omega_{AB}]. \label{eq:linearBound}
\end{align}
Inserting this relation in \eqref{eq: perp term obj function second} leads to an objective function expressed only in terms of finite-dimensional quantities. The full optimization problem, including the constraints found in Appendix \ref{app:DimRed}, is written as
\begin{equation} \label{eq:lower bound optimization}
\begin{gathered}
    \inf_{\tilde\omega, q, s, t} 
    \frac{\alpha}{\alpha-1} \Bigg[\sum_{c \in \tilde{\mc{C}}} q(c) \log \left( \frac{q(c)}{\bar{p}(c)(\Tr[ \hat{\Pi}^c\tilde\omega] +  s(c) + 2 t(c))} \right) 
    + q(\perp) \log \left( \frac{q(\perp)}{\pkey(\hat\Psi_\mu(\tilde\omega)- \Tr[R^\top \tilde\omega] + \tilde\zeta_\alpha(k))}\right)
    \Bigg] 
    \\
    \mathrm{s.t.} \quad  \tilde\omega_{AB} \succeq 0, \;\sigma_A - \Tr_B[\tilde\omega_{AB}] \succeq 0, \,
    \\
      \sum_{c\in\mc C} q(c)=1, \; q(c)\geq 0, \; c\in\mc C, 
    \\
    % k \leq \frac{2\sum_{z=4}^7 \sqrt{ (\sum_{x=0}^3 \Tr[\hat{\Pi}^{(x,z)}\tilde{\omega}])(\sum_{x=0}^3 s(x,z))}}{\Tr[\g (\tilde{\omega}) + s(\bot)]}, \\
    s(c)\geq 0 \quad \forall c\in \tilde{\mc C}, \; s(\perp) = \sum_{x=0}^3\sum_{z=4}^7 s(x,z), \\
    \sum_{z=0}^8 s(x,z) = \langle x|\sigma_A-\mathrm{Tr}_B[\tilde\omega]|x\rangle  \quad \forall x , \\
    \bra{N_c+1}R_B^\top\ket{N_c+1}\sum_{z=0}^3s(x,z) \leq
 \left(\sum_{z=0}^3\bra{N_c+1}\tilde R_B^z\ket{N_c+1}\right)
 \sum_{z=4}^7s(x,z),\quad\forall x,
 \\
\bra{N_c+1}\tilde R_B^8\ket{N_c+1}\sum_{z=4}^7s(x,z) \leq
 \bra{N_c+1}R_B^\top\ket{N_c+1}\,s(x,8),\quad\forall x,
 \\
         t(c)^2 \leq \Tr[\hat\Pi^c\tilde\omega]\, s(c)   \quad \forall c \in \mc{C},\\
    t(x,8)=\sum_{z=0}^3 t(x,z) = \sum_{z=4}^7 t(x,z) = 0 \quad \forall x,
\end{gathered}
\end{equation} 
where $k$ is defined in Eq. \eqref{eq: final penalization argument}.
\subsection{Explicit conic formulation}
To rewrite the optimization problem in \eqref{eq:lower bound optimization} as a conic program, we make use of the following cones \cite{coey2022solving, navarro2025}:
\begin{align}
    \mc{K}_\mathrm{RE} &= \cl\left\{(u,p,q) \in \R \times \R_>^d \times \R_>^d ; u \geq \sum_{i=1}^d q(i) \log[q(i)/p(i)]\right\}, 
    \\
    \mc{K}^p_\mathrm{pc} &= \cl\{(x,y,m) \in \R_\ge^2 \times \R : x^p y^{(1-p)} \geq |m|\},
       \\
    \mc{K}_\mathrm{rs} &= \cl\{(x,y,z)\in \R_\ge \times \R_\ge \times \R : 2 x y \geq z^2 \},
        \\
    \mc{K}^{\mu, \gmap, \zgmap, S}_\text{FastRényiQKD} &= \cl\left\{(u,\rho) \in \R \times \mathbb H^{4(N_c+1)}_\succ; u \ge - \Psihat(\rho) \right\}\,,
\end{align}
which respectively represent the vector relative entropy cone, the power cone, the rotated second-order cone, and the FastRényiQKD cone (after applying the corresponding facial reduction). Thus,
\begin{equation} \label{eq: main optimization problem 2}
\begin{gathered}
    h_\alpha 
    \geq 
    \inf_{\tilde\omega, q, s, t, u, g, v, w, h_\mathrm{RE}, h_\mathrm{QKD}} 
    \frac{\alpha}{\alpha-1} \Bigg[h_\mathrm{RE}
    + h_\mathrm{QKD}
    \Bigg] 
    \\
    \mathrm{s.t.} \quad \sigma_A - \Tr_B[\tilde\omega_{AB}] \succeq 0, \; \sum_{c\in\mc C} q(c)=1,
    \\
    \left(h_\mathrm{RE} ,\left(\bar{p}(c)[\Tr[ \hat{\Pi}^c\tilde\omega] +  s(c) + 2 t(c) ],q(c)_{c\in \tilde{\mc C}}\right)_{c\in \tilde{\mc C}}\right)\in \mc{K}_\mathrm{RE}
    \\
     \left(h_\mathrm{QKD},\pkey(v- \Tr[R^\top \tilde\omega] -  u ),q(\perp) \right)\in \mc{K}_\mathrm{RE}
        \\
{\left(\frac{1}{2} \Tr[\hat\Pi^{(x,z)} \tilde\omega ],s(x,z),g(x,z)\right)\in\mc K_\mathrm{rs}, \; x \in \{0,\ldots,3\} , \; z \in \{4,\ldots,7\},}
\\
\left(w, \Tr[R^\top \tilde\omega]+ s(\perp)+ g_\Sigma , g_\Sigma\right)\in \mc K^{1/\alpha}_{pc} 
\\
    \left(\Tr[R^{\top}\tilde\omega] + s(\perp) + g_\Sigma |\hat{\mc{Z}}|^{\alpha-1} - w, \Tr[R^\top \tilde\omega]+ s(\perp) + g_\Sigma, v \right)\in \mc K^{1/\alpha}_{pc}
    \\
    g(x,z)\geq 0,
    \\
    (u, \tilde\omega) \in \mc{K}^{\mu, \gmap, \zgmap, S}_\text{FastRényiQKD} 
    \\
        \sum_{z=0}^8 s(x,z) = \langle x|\sigma_A-\mathrm{Tr}_B[\tilde\omega]|x\rangle  \quad \forall x, 
    \\
    \bra{N_c+1}R_B^\top\ket{N_c+1}\sum_{z=0}^3s(x,z) \leq
 \left(\sum_{z=0}^3\bra{N_c+1}\tilde R_B^z\ket{N_c+1}\right)
 \sum_{z=4}^7s(x,z),\quad \forall x,
 \\
 \bra{N_c+1}\tilde R_B^8\ket{N_c+1}\sum_{z=4}^7s(x,z) \leq
 {\bra{N_c+1}R_B^\top\ket{N_c+1}\,s(x,8),\quad \forall x,}
    \\
    \left(\frac{1}{2} \Tr[\hat\Pi^{(x,z)} \tilde\omega ], s(x,z), t(x,z)\right) \in \mc{K}_{rs}  \quad x \in \{0,\ldots,3\}, \; z \in \{0,\ldots,7\},\\
    t(x,8)=\sum_{z=0}^3 t(x,z) = \sum_{z=4}^7 t(x,z) = 0 \quad \forall x.\\
\end{gathered}
\end{equation} 
where $g_\Sigma = \sum_{x=0}^3 \sum_{z=4}^7 g(x,z)$ and $s(\perp) = \sum_{x=0}^3\sum_{z=4}^7 s(x,z)$.

\section{Variable-length secret key rates}\label{App:VarLengthKeys}
Let us now provide our framework for the derivation of secret keys under a MEAT-based, variable-length approach. This section computes the secret key length in accordance with the formula \cite{kamin25MEATsecurity} 
\begin{align}
\ell \leq n\bigl(f\cdot \mathrm{F}^\mathrm{obs}_{c_1^n}+\kappa\bigr)
-\mathrm{leak}(c_1^n)
-\frac{\alpha}{\alpha-1}
\log\frac1{\varepsilon_{\mathrm{PA}}}
+2.
\end{align}

The tradeoff function $f$ can be arbitrarily chosen, but its optimal value is found by solving the program
\begin{equation}
\begin{gathered}     \inf_{{\lambda}\in S_{\Omega}} \inf_{\omega \in \Sigma_\Omega^{(m)}} \frac{\alpha}{\alpha-1} D_\mathrm{KL}(\lambda\|\mc{M}(\omega_{AB})_{C}) + \lambda(\perp ) H^\uparrow_\alpha (Z|E)_{\widetilde{\mc{M}} (\omega)_{|\perp}}, \\
\text{s.t. }\Tr_{B}[\omega_{AB}]=\sigma_{A}, \; \omega_{AB} \succeq 0 \\
\lambda - p = 0
\end{gathered}
\end{equation}
Namely, the fixed-length program conditioned only on the reference probability $p$. Then, taking the dual value corresponding to the last constraint guarantees an optimal choice for the tradeoff \cite{kamin25MEATsecurity}. Hence, we may find such a function by adding the slack $\lambda$ in the optimization described in Appendix \ref{app:FixedLength}. Once a value for $f$ is set, the coefficient $\kappa$ results from the optimization \eqref{eq:kappaInitial} over the corresponding $f$-weighted Rényi entropy \cite{arqand2024generalized,arqand2025MEAT} defined as
\begin{equation}\label{eq:fWeightedRenyi}
     H^{\uparrow,f}_\alpha (Z|CE)_{\mathcal{M}(\omega)}= \frac{\alpha}{1-\alpha} \log \left( \sum_{{c}\in \mc{C}} p({c}) 2^{\frac{1-\alpha}{\alpha} (-f ({c})+ H^\uparrow_\alpha (Z|E))_{{\mathcal{M}(\omega)}_{|{c}}}}  \right).
\end{equation}
Using the remark in \cite{navarro2025}, this entropy can be lower bounded as 
\begin{align}
    H^{\uparrow,f}_\alpha (Z|{C}E)_{\tilde{\mc{M}}(\omega)} &= \frac{\alpha}{1-\alpha} \log \Bigg( \sum_{c \in \tilde{\mc{C}}} \bar{p}(c) \Tr[{\mc{M}}(\omega)_{|c}] 2^{\frac{\alpha-1}{\alpha}f (c)}   \nonumber \\  
    & \quad +\bar{p}(\bot) \Tr[\mc{M}(\omega)_{|\bot}] 2^{\frac{1-\alpha}{\alpha} (-f ({\perp})+ H^\uparrow_\alpha (Z|E)_{\tilde{\mc{M}}(\omega)_{|{\perp}}})}  \Bigg) % \nonumber\\
\end{align}
We identify the components in $C$ using \eqref{eq:OneRoundCh} to find their finite-dimensional representation, as well as \eqref{eq:KeyMapDecomp}. Furthermore, as the entropies corresponding to $c \in \tilde{\mc{C}}$ deterministically generate $Z=\bot$, so they are equal to zero. Next, we ought to make use of the continuity bounds to pinch the state $\omega$ into a block-diagonal structure. Combining this result with the lower bound on the sandwiched Rényi entropy \cite{navarro2025}, Proposition \ref{prop:renyi continuity bound} and the simplifications at \eqref{eq:infinityBound}
\begin{align}
    H^\uparrow_\alpha (Z|E)_{\widetilde{\mc{M}} (\omega)_\perp} 
    &\geq -\frac{\alpha}{\alpha-1} \log \left( \frac{ \Psi_\mu\big(\g(\hat\omega),\zg(\hat\omega)\big) }{\Tr[\g({\omega})]} \right)
    - \zeta_\alpha \left(k \right) \nonumber\\
    &= -\frac{\alpha}{\alpha-1} \log \left( \frac{\Psi_\mu\big(\g(\hat\omega),\zg(\hat\omega)\big)}{\Tr[\g({\omega})] 2^{\frac{1-\alpha}{\alpha}\zeta_\alpha(k)}} \right) \nonumber \\
    &\geq -\frac{\alpha}{\alpha-1} \log \left( \frac{\hat{\Psi}_\mu(\tilde{\omega}) + s(\bot)}{\Tr[\g({\omega})] 2^{\frac{1-\alpha}{\alpha}\zeta_\alpha(k)}} \right).
\end{align}
Applying all the derivations corresponding to \eqref{eq: simplified penalization} and \eqref{eq:linearBound} we eventually arrive at
\begin{align}
     H^\uparrow_\alpha (Z|E)_{\widetilde{\mc{M}} (\omega)_\perp} 
    \geq  - \frac{\alpha}{\alpha-1} \log \left( \frac{\hat{\Psi}_\mu(\tilde{\omega}) + \tilde{\zeta}_\alpha(k) - \Tr[R^\top \tilde{\omega}] }{\Tr[R^\top {\omega}]}\right)
\end{align}
Inserting this relation in the expression for the $f-$weighted Rényi entropy, and taking \eqref{eq:OneRoundCh}
\begin{align}
    H^{\uparrow,f}_\alpha (Z|{C}E)_{\widetilde{\mc{M}}(\omega)} \geq \frac{\alpha}{1-\alpha}& \log \Bigg( \sum_{c \in \tilde{\mc{C}}} \bar{p}(c) [\Tr[{\hat{\Pi}^c}\tilde{\omega}] + s(c) + 2t(c)] 2^{\frac{\alpha-1}{\alpha}f (c)} \nonumber \\
    & \quad+  \pkey 2^{\frac{\alpha-1}{\alpha} f ({\perp})} (\hat{\Psi}_\mu(\tilde{\omega}) + \tilde{\zeta}_\alpha(k) - \Tr[R^\top \tilde{\omega}] ) \Bigg). \label{eq:fWeightedReduced}
\end{align}

\subsection{Explicit conic formulation} \label{App:ConicKappa}

The right-hand side of \eqref{eq:fWeightedReduced} is expressed in terms of finite-dimensional elements, allowing the calculation of a lower bound on $\kappa$ by minimizing the right-hand side in accordance with \eqref{eq:kappaInitial}. The feasible set corresponding to this minimization can be adapted thanks to the reductions accordingly explained in Appendix \ref{app:DimRed}. As the negative logarithm is monotonically decreasing, we can perform the optimization by simply maximizing the argument. Furthermore, we have that the original feasible set is identical to the one provided for the fixed-length implementation, such that we formulate the maximization according to all the conic formulations explained for the fixed-length implementation (cf. Appendix \ref{app:FixedLength}).

As a result, we arrive at
\begin{equation}\label{eq:KappaConic}
\begin{gathered}
    \max_{\tilde\omega, s, t, u, g, v, w} 
    \sum_{c \in \tilde{\mc{C}}} \bar{p}(c) [\Tr[ \hat{\Pi}^c\tilde\omega] +  s(c) + 2 t(c)] 2^{\frac{\alpha-1}{\alpha}f (c)}  + \pkey 2^{\frac{\alpha-1}{\alpha} f ({\perp})} (-u + v - \Tr[R^\top \tilde{\omega}] ) 
    \\
    \mathrm{s.t.} \quad \sigma_A - \Tr_B[\tilde\omega_{AB}] \succeq 0,
    \\
\left(\frac{1}{2} \Tr[\hat\Pi^{(x,z)} \tilde\omega ],s(x,z),g(x,z)\right)\in\mc K_\mathrm{rs}, \; x \in \{0,\ldots,3\} , \; z \in \{4,\ldots,7\},
\\
\left(w, \Tr[R^\top \tilde\omega]+ s(\perp)+ g_\Sigma , g_\Sigma\right)\in \mc K^{1/\alpha}_{pc},
\\
    \left(\Tr[R^{\top}\tilde\omega] + s(\perp) + g_\Sigma |\hat{\mc{Z}}|^{\alpha-1} - w, \Tr[R^\top \tilde\omega]+ s(\perp) + g_\Sigma, v \right)\in \mc K^{1/\alpha}_{pc},
    \\
    g(x,z)\geq 0,
    \\
    (u, \tilde\omega) \in \mc{K}^{\mu, \gmap, \zgmap, S}_\text{FastRényiQKD},
    \\
    \sum_{z=0}^8 s(x,z) = \langle x|\sigma_A-\mathrm{Tr}_B[\tilde\omega]|x\rangle  \quad x \in \{0,\ldots,3\} , 
    \\
    \bra{N_c+1}R_B^\top\ket{N_c+1}\sum_{z=0}^3s(x,z) \leq
 \left(\sum_{z=0}^3\bra{N_c+1}\tilde R_B^z\ket{N_c+1}\right)
 \sum_{z=4}^7s(x,z),\quad x \in \{0,...,3\},
 \\
 \bra{N_c+1}\tilde R_B^8\ket{N_c+1}\sum_{z=4}^7s(x,z) \leq
 \bra{N_c+1}R_B^\top\ket{N_c+1}\,s(x,8),\quad  x \in \{0,\ldots,3\},
    \\
        \left(\frac{1}{2} \Tr[\hat\Pi^{(x,z)} \tilde\omega ], s(x,z), t(x,z)\right) \in \mc{K}_{rs}  \quad  x \in \{0,\ldots,3\}, \; z \in \{0,\ldots,7\},\\
    t(x,8)=\sum_{z=0}^3 t(x,z) = \sum_{z=4}^7 t(x,z) = 0 \quad  x \in \{0,\ldots,3\}.
\end{gathered}
\end{equation} 
where $g_\Sigma = \sum_{x=0}^3 \sum_{z=4}^7 g(x,z)$ and $s(\perp) = \sum_{x=0}^3\sum_{z=4}^7 s(x,z)$.

Since the constraints on the contributions above the cutoff are valid for every $m>N_c$, the above resulting lower bound on $\kappa$ is independent of $m$.

\section{Proof of Theorem \ref{Th:MainTheorem}} \label{app:SecurityProof}

Using the security proof developed in Ref.  \cite{kamin25MEATsecurity}, we first establish security for an arbitrary finite-dimensional protocol with $m>N_c$. Throughout the following calculations, we omit the superscript $(m)$ on the corresponding states for simplicity. The extension to the physical infinite-dimensional state is obtained at the end of the proof.

Starting from the definition of security for QKD in Eq. \eqref{eq:secureVarLen}, we split the trace norm into the correctness and secrecy clauses by introducing the auxiliary state \(\widehat\rho_{K_AK_BFE'L}\), obtained from \(\rho_{K_AK_BFE'L}\), by replacing Alice’s classical key with a copy of Bob’s \cite[Theorem 4.1]{portmann2014cryptographicsecurityquantumkey}. Using the triangle inequality for each event, we have 
\begin{align}
    &T(\rho_{K_AK_B F E' L},\bar{\tau}_{K_AK_B F E' L}) \nonumber \\ 
& \leq T(\rho_{K_AK_B FE' L},\widehat\rho_{K_AK_BFE'L}) + T(\widehat\rho_{K_AK_BFE'L},\bar{\tau}_{K_AK_B F E' L}) \nonumber \\
& = \sum_\ell\Pr_\rho[\Omega_\ell] T(\rho_{K_AK_B FE' L|\Omega_\ell},\widehat\rho_{K_AK_BFE'L|\Omega_\ell)} + T(\widehat\rho_{K_AK_BFE'L},\bar{\tau}_{K_AK_B F E' L}) 
\nonumber \\
& \leq \sum_\ell\Pr_\rho[\{ K_A \neq K_B \} \wedge \Omega_\ell] + T(\rho_{K_B  F E' L},\bar{\tau}_{K_B F E' L}). \label{eq:almostCorrect}
\end{align}
In the third line, we split the first trace according to a successful key generation $\Omega_\ell$, (which particularly includes a successful error verification $\Omega_\mathrm{EV}$), and a failed one $ \Omega_0$ (where the ideal state is reached, but no secret key is generated). Accordingly, we replace the remaining trace in the fourth line; and since Alice’s key is a copy of Bob’s in both the auxiliary and ideal states, tracing out \(K_A\) in the last line leaves the trace distance unchanged.

For the correctness, we use the definition of universal$_2$ hash to find, for a validation function $g(\cdot)$ of length $\lceil \log(1/\varepsilon_\mathrm{EV}) \rceil$ \cite{nahar2024postselection_p&mcvqkd} 

\begin{align}
    \sum_\ell\Pr[\{ K_A \neq K_B \} \wedge \Omega_\ell]  &\leq \Pr[\{Z_1^n \neq \hat{Z}_1^n \} \wedge \Omega_\mathrm{EV}]\nonumber \\
    &= \Pr[\{Z_1^n \neq \hat{Z}_1^n \} \wedge \{g(Z_1^n) = g(\hat{Z}_1^n)\}] \nonumber\\
    &\leq \Pr[ g(Z_1^n) = g(\hat{Z}_1^n) | Z_1^n \neq \hat{Z}_1^n]\nonumber \\
    &\leq 2^{-\lceil \log(1/\varepsilon_\mathrm{EV}) \rceil} \leq \varepsilon_\mathrm{EV}. \label{eq:CorrectnesViaHash}
\end{align}
In the first step we apply the data processing inequality for the privacy amplification hash, then we explicitly replaced $\Omega_\mathrm{EV}$ by the comparison of the hashes. The third line follows from Bayes' theorem, and in the fourth line we used the definition of collision probability for universal$_2$ hash functions. As a result, we replace in \eqref{eq:almostCorrect}
\begin{align}
    T(\rho_{K_AK_B F E'L},\bar{\tau}_{K_AK_B F E'L}) &\leq  \varepsilon_\mathrm{EV} + T(\rho_{K_BFE' L},\bar{\tau}_{K_BFE' L}),
\label{eq:LastTrace}
\end{align}
Next we reduce the final trace to achieve secrecy. We recall $\mathrm{Pr}_\rho [\Omega_\ell]$ as the probability of the event $\Omega_\ell$, where Alice and Bob validate their keys and distill a secret key of length $\ell$. Then we decompose the trace distance between the original and target states according to the different possible lengths,

\begin{align}\label{eq:SumProbs}
    T(\rho_{K_BFE' L},\bar{\tau}_{K_BFE' L}) = \sum_{\ell=0}^\infty \mathrm{Pr}_{\rho}[\Omega_\ell] 
    T( \rho_{K_B FE'L|\Omega_\ell},\bar{\tau}_{K_B FE'L|\Omega_\ell}).
\end{align}
Let \(\ell(c_1^n)\) denote the secret key length selected according to \(\eqref{eq: final key}\) for the observed announcements \(C_1^n=c_1^n\). Since \(C_1^n\) is classical and included in \(E'\), we may further decompose each event \(\Omega_\ell\) according to the observed announcement strings \(c_1^n\) and a successful verification. We then apply the Rényi leftover hash lemma, as described in Appendix \ref{App:PrivacyAmp}, and the bound \eqref{eq:LeakRemoval} for each term in the sum
\begin{align}\label{eq:SumProbs1}
    T(\rho_{K_BFE' L},\bar{\tau}_{K_BFE' L}) 
    \leq 
    \sum_{\{c_1^n | \ell(c_1^n)>0 \}} \mathrm{Pr}_\rho[\Omega_{c_1^n} \wedge \Omega_\mathrm{EV}]  
    2^{\frac{\alpha-1}{\alpha}[\ell(c_1^n)-H^\uparrow_\alpha(Z_1^n| E')_{\rho_{|\Omega_{c_1^n} \wedge \Omega_\mathrm{EV}}} + \mathrm{leak}(c_1^n) -2]}.
\end{align}
Now we apply the hypothesis \eqref{eq: final key} to the key length size, obtaining
\begin{subequations}\label{eq:SumProbs2}
\begin{align}
    \sum_{\{c_1^n | \ell(c_1^n)>0 \}} &\mathrm{Pr}_\rho[\Omega_{c_1^n} \wedge \Omega_\mathrm{EV}]   2^{\frac{\alpha-1}{\alpha} \left[ \hat{f}(c_1^n) -H^\uparrow_\alpha(Z_1^n| E')_{\rho_{|\Omega_{c_1^n} \wedge \Omega_\mathrm{EV}}}\right] - \log\frac1{\varepsilon_{\mathrm{PA}}} }  \nonumber \\
    &= \varepsilon_{\mathrm{PA}} \sum_{\{c_1^n | \ell(c_1^n)>0 \}} \mathrm{Pr}_\rho[\Omega_{c_1^n} \wedge \Omega_\mathrm{EV}]   2^{\frac{\alpha-1}{\alpha} \left[ \hat{f}(c_1^n) -H^\uparrow_\alpha(Z_1^n| E')_{\rho_{|\Omega_{c_1^n} \wedge \Omega_\mathrm{EV}}}\right]}
    \label{subeq: secrecy 2}  \\
     & \leq \varepsilon_{\mathrm{PA}} \sum_{\{c_1^n | \ell(c_1^n)>0 \}} \mathrm{Pr}_\rho[\Omega_{c_1^n}]   2^{\frac{\alpha-1}{\alpha} \left[ \hat{f}(c_1^n) -H^\uparrow_\alpha(Z_1^n| E')_{\rho_{|\Omega_{c_1^n} }}\right]}
    \label{subeq: secrecy 3}  \\
    &\leq \varepsilon_{\mathrm{PA}} \sum_{\{c_1^n \in\mc{C}^n \}} \mathrm{Pr}_\rho[\Omega_{c_1^n}]   2^{\frac{\alpha-1}{\alpha} \left[ \hat{f}(c_1^n) -H^\uparrow_\alpha(Z_1^n| E')_{\rho_{|\Omega_{c_1^n} }}\right]}\label{subeq: secrecy 4}  \\
    &= \varepsilon_{\mathrm{PA}} 2^{- \frac{\alpha-1}{\alpha} H_\alpha^{\uparrow,\hat{f}}(Z_1^n|E')_\rho} \label{subeq: secrecy 5}  \\
    &\leq \varepsilon_{\mathrm{PA}}. \label{subeq: secrecy 6} 
\end{align}
\end{subequations}
In line \eqref{subeq: secrecy 3}, we use \cite[Lemma B.5]{dupuis2016entropy} to remove the conditioning on $\Omega_\mathrm{EV}$, where the resulting correction cancels the corresponding conditional probability in the preceding expression. In line \eqref{subeq: secrecy 4}, we extend the sum to all possible announcement strings $c_1^n$, which can only increase the expression. Then, line \eqref{subeq: secrecy 5} follows from the definition of the $\hat f$-weighted Rényi entropy. Finally, after removing the conditioning on $\Omega_{\mathrm{EV}}$,
the register $C_{\mathrm{EV}}$ is independent of $Z_1^n\hat E$
conditional on $C_1^n$. Conditioning on $C_{\mathrm{EV}}$
therefore does not change the $\hat f$-weighted entropy,
and \eqref{subeq: secrecy 6} follows from
Proposition~\ref{prop:MEAT}. Altogether, this results in the desired $\varepsilon_\mathrm{PA}-$secrecy. Identifying this last result with \eqref{eq:LastTrace}, we finally find that the finite-dimensional protocol is $\varepsilon_\mathrm{EV}-$correct and $\varepsilon_\mathrm{PA}-$secret
\begin{equation} \label{eq: security finite dim}
   T(\rho_{K_AK_B F E'L},\bar{\tau}_{K_AK_B F E' L}) \leq  \varepsilon_\mathrm{EV} + \varepsilon_\mathrm{PA}.
\end{equation}
To extend this security statement to the infinite-dimensional protocol, let \(\bar\omega\) be an arbitrary physical state and \(\bar\omega^{(m)}\) its finite-dimensional approximations constructed in Lemma \ref{lem:MEAT marginal approximation}. We denote by $\rho^{(\infty)}$ and $\bar\tau^{(\infty)}$ the corresponding real and ideal output states of the physical protocol, respectively, whereas $\rho$ and $\bar\tau$ denote the outputs corresponding to $\bar\omega^{(m)}$, as considered in the finite-dimensional proof above. Therefore,
\begin{align}
T\left(\rho_{K_AK_B F E'L}^{(\infty)},\bar\tau_{K_AK_B F E'L}^{(\infty)}\right)
&\leq 
T(\rho_{K_AK_B F E'L}^{(\infty)},\rho_{K_AK_B F E'L})   + T(\bar\tau_{K_AK_B F E'L},\bar\tau_{K_AK_B F E'L}^{(\infty)})
\nonumber\\
& \qquad  \qquad\qquad+ T\left(\rho_{K_AK_B F E'L},\bar\tau_{K_AK_B F E'L}\right) 
\nonumber\\
&\leq
\varepsilon_{\mathrm{EV}}+\varepsilon_{\mathrm{PA}}
+\left\|\bar\omega-\bar\omega^{(m)}\right\|_1 
,
\end{align}
where in the first line we used the triangle inequality, while in the second line we applied the bound in Eq. \eqref{eq: security finite dim} and the data processing inequality. Applying Lemma \ref{lem:MEAT marginal approximation} by taking $m\to \infty$, we conclude

\begin{equation}
    T\left(\rho_{K_AK_B F E'L}^{(\infty)},\bar\tau_{K_AK_B F E'L}^{(\infty)}\right) \leq \varepsilon_{\mathrm{EV}}+\varepsilon_{\mathrm{PA}}.
\end{equation}
Thus, the security bound established for arbitrary finite-dimensional approximations extends to the physical infinite-dimensional optical protocol.

\section{Hyperplane approximation of the Rényi cone}\label{app:Hyperplanes}

Starting from the conic formulation \eqref{eq:KappaConic}, we obtain a more efficiently computable lower bound to the key rate by approximating the FastRényiQKD cone by a series of hyperplanes. Let $\{\sigma_i\}_{i\in \mc{I}}$ be a set of positive definite Hermitian matrices, and $\nabla \Psihat(\sigma_i)$ the gradient of $\Psihat$ at the point $\sigma_i$. Since $-\Psihat$ is a convex function, at any point it is lower-bounded by its linearization, such that
\begin{equation}
-\Psihat(\tilde{\omega}_{AB}) \ge -\Psihat(\sigma_i) - \langle \nabla \Psihat(\sigma_i), \tilde{\omega}_{AB} - \sigma_i\rangle, \quad\forall i \in \mc{I},
\end{equation}
which means that this set of inequalities forms an outer approximation of the cone $\mc{K}^{\mu, \gmap, \zgmap, S}_\text{FastRényiQKD}$. This implies that we can lower-bound the optimal value of problem \eqref{eq:KappaConic} by applying the replacement
\begin{align}\label{eq:replacement}
    (u, \tilde\omega) \in \mc{K}^{\mu, \gmap, \zgmap, S}_\text{FastRényiQKD} \to u \ge -\Psihat(\sigma_i) - \langle \nabla \Psihat(\sigma_i), \tilde{\omega}_{AB} - \sigma_i\rangle, \quad\forall i \in \mc{I}
\end{align}
Additionally, this requires adding the previously implicit constraint $\tilde{\omega}_{AB} \succeq 0$ to make it a better approximation. The advantage of this bound is that it is much easier to compute, and is equal to the optimal value of problem \eqref{eq:KappaConic} if $\{\sigma_i\}_{i \in \mc{I}}$ contains the optimal point.

\subsection{Construction of the hyperplane set}

The previous discussion relied on the existence of a set $\{\sigma_i\}_{i \in \mc{I}}$ of  $|\mc{I}|$ states that define a suitable approximation for the FastRényiQKD cone via hyperplanes. Let us now illustrate an efficient method to build said set, as well as finding points that are close to the actual, optimal value while exploiting a reduced dimensionality.

The fundamental observation is that the optimal solutions have very small amplitudes for large numbers of photons on Bob's side. We expect then to provide a good approximation by solving the original problem with a smaller cutoff $N_c \le N_c'$, obtaining an optimal solution $\sigma^{(N_c)}_{AB}$, and padding it with zeros on Bob's side. We then find a point in the larger space
\begin{align}
    \tilde \sigma^{(N'_c)}_{AB} := \sigma^{(N_c)}_{AB} \oplus (\id_A \otimes0_B^{(N'_c - N_c)}).
\end{align}
We emphasize that $\tilde \sigma^{(N'_c)}_{AB}$ does not need to be close to the optimal solution or even a feasible solution for the hyperplane approximation to give a valid lower bound to the key rate, as the derivation above is valid for any set of positive definite matrices. The only consequence of using a $\tilde \sigma^{(N'_c)}_{AB}$ far from the optimal solution is a loose bound on the key rate.

We need nevertheless positive definiteness, and $\tilde \sigma^{(N'_c)}_{AB}$ has by construction null eigenvalues. To overcome this inconvenience, we define its depolarized counterpart

\begin{align}
    \sigma_0 := (1- 4(N_c+1)\epsilon)\tilde \sigma^{(N'_c)}_{AB} + 4(N_c + 1) \epsilon \frac{\id_{A}\otimes \Pi^{(N'_c)}_B }{4(N'_c + 1)}
\end{align}
for some $\epsilon > 0$, so that its smallest eigenvalues are equal to $\epsilon$. Numerical experiments show that $\epsilon \approx 10^{-13}$ is enough to achieve numerical stability.

This provides the first point $\sigma_0$ in the set $\{\sigma_i\}_{i\in \mc{I}}$ needed to construct the hyperplane approximation. To obtain the subsequent points, we iteratively refine it by solving problem \eqref{eq:KappaConic} with the replacement \eqref{eq:replacement} with the existing points, and expand the set by adding the optimal solution just found, depolarizing as needed.

\subsection{Computation of the dual problem}\label{sec:dual}
 Let us now compute the dual of problem \eqref{eq:KappaConic} with the replacement \eqref{eq:replacement}, which can be used to produce certified lower bounds. For that, we build the Lagrangian
 \begin{multline}
	 L(\tilde{\omega},s,t,u,g,v,w,\omega^*,\tau_A,a_1, a_2, a_3, b_1, b_2, b_3, c_1, c_2, c_3, g^*, \gamma, \lambda, \mu, \nu, d_1, d_2, d_3, t_8^*, t_{03}^*, t_{47}^*) =\\
\sum_{c \in \tilde{\mc{C}}} \bar{p}(c) [\Tr[ \hat{\Pi}^c\tilde\omega] +  s(c) + 2 t(c)] 2^{\frac{\alpha-1}{\alpha}f (c)}  + \pkey 2^{\frac{\alpha-1}{\alpha} f ({\perp})} (-u + v - \Tr[R^\top \tilde{\omega}] ) \\
+ \langle \omega^*, \omega \rangle \\
+ \langle \tau_A, \sigma_A - \Tr_B[\tilde\omega] \rangle \\
+ \sum_{x =0}^3\sum_{z=4}^7\left(a_1(x,z) \frac{1}{2} \Tr[\hat\Pi^{(x,z)} \tilde\omega ] + a_2(x,z)s(x,z) + a_3(x,z)g(x,z)\right) \\
+ \left(b_1 w + b_2(\Tr[R^\top \tilde\omega]+ s(\perp)+ g_\Sigma) + b_3 g_\Sigma\right)\\
  + \left(c_1(\Tr[R^{\top}\tilde\omega] + s(\perp) + g_\Sigma |\hat{\mc{Z}}|^{\alpha-1} - w) + c_2( \Tr[R^\top \tilde\omega]+ s(\perp) + g_\Sigma) + c_3 v \right)\\
  + \langle g^*, g\rangle \\
  + \sum_i \gamma_i (u  + \Psihat(\sigma_i) + \langle \nabla \Psihat(\sigma_i), \tilde{\omega} - \sigma_i\rangle) \\
+ \sum_x \lambda_x \left(\langle x|\sigma_A-\mathrm{Tr}_B[\tilde\omega]|x\rangle -  \sum_{z=0}^8 s(x,z)\right) \\
+\sum_x \mu_x\left[\left(\sum_{z=0}^3\bra{N_c+1}\tilde R_B^z\ket{N_c+1}\right)\sum_{z=4}^7s(x,z) - \bra{N_c+1}R_B^\top\ket{N_c+1}\sum_{z=0}^3s(x,z)\right] \\
+\sum_x \nu_x\left[\bra{N_c+1}R_B^\top\ket{N_c+1}\,s(x,8) - \bra{N_c+1}\tilde R_B^8\ket{N_c+1}\sum_{z=4}^7s(x,z)\right] \\
+\sum_{x=0}^3\sum_{z=0}^7\left(d_1(x,z)\frac{1}{2} \Tr[\hat\Pi^{(x,z)} \tilde\omega ] + d_2(x,z) s(x,z) + d_3(x,z) t(x,z)\right) \\
+ \sum_x \left(t_8^*(x) t(x,8) + t_{03}^*(x)\sum_{z=0}^3 t(x,z) + t_{47}^*(x)\sum_{z=4}^7 t(x,z)\right)
\end{multline}
where $\omega^* \succeq 0$, $\tau_A \succeq 0$, $\left(a_1(x,z),a_2(x,z),a_3(x,z)\right) \in\mc K_\mathrm{rs}, \; x \in \{0,\ldots,3\} , \; z \in \{4,\ldots,7\}$, $\left(b_1, b_2, b_3 \right)\in \mc{K}^{*,1/\alpha}_{pc}$, $\left(c_1, c_2, c_3\right)\in \mc K^{*,1/\alpha}_{pc}$, $g^* \ge 0$, $\gamma \ge 0$, $\mu \ge 0$, $\nu \ge 0$, $(d_1(x,z), d_2(x,z), d_3(x,z) \in \mc{K}_{rs}, x \in \{0,\ldots,3\} , \; z \in \{0,\ldots,7\}$.

We now isolate the original variables
 \begin{multline}
	 L=\Big\langle \sum_{c \in \tilde{\mc{C}}} \bar{p}(c) 2^{\frac{\alpha-1}{\alpha}f (c)}\hat{\Pi}^c - \pkey 2^{\frac{\alpha-1}{\alpha} f ({\perp})} R^\top - \tau_A \otimes \id_B + \sum_{x =0}^3\sum_{z=4}^7a_1(x,z) \frac{1}{2} \hat\Pi^{(x,z)} + (b_2 + c_1 + c_2) R^\top \\ 
     + \sum_i \gamma_i \nabla\Psihat(\sigma_i) - \sum_x \lambda_x \proj{x} \otimes \id_B + \sum_{x=0}^3\sum_{z=0}^7d_1(x,z)\frac{1}{2} \hat\Pi^{(x,z)} + \omega^*, \tilde{\omega}\Big\rangle \\
	 +(\sum_i \gamma_i -\pkey 2^{\frac{\alpha-1}{\alpha} f ({\perp})}) u \\
	 +(c_3 + \pkey 2^{\frac{\alpha-1}{\alpha} f ({\perp})})  v \\
	 +(b_1-c_1)w \\
	 + \langle g^* + a_3 + (b_2 + b_3 + c_1|\hat{\mc{Z}}|^{\alpha-1} + c_2) \mathbf{1}, g\rangle \\
	 +\Big\langle \bar{p} \odot 2^{\frac{\alpha-1}{\alpha}f} + a_2\odot\mathbf{1}_{03,47} +d_2\odot\mathbf{1}_{03,07} + (b_2 + c_1 + c_2)\mathbf{1}_{03,47}- \ket{\lambda}\bra{\mathbf{1}_{08}} \\ 
	 \left(\sum_{z=0}^3\bra{N_c+1}\tilde R_B^z\ket{N_c+1}\right)\ket{\mu}\bra{\mathbf{1}_{47}} -\bra{N_c+1}R_B^\top\ket{N_c+1}\ket{\mu}\bra{\mathbf{1}_{03}} \\
\bra{N_c+1}R_B^\top\ket{N_c+1}\ket{\nu}\bra{\mathbf{1}_8} - \bra{N_c+1}\tilde R_B^8\ket{N_c+1} \ket{\nu}\bra{\mathbf{1}_{47}},  s \Big\rangle \\
+\Big\langle 2\bar{p} \odot 2^{\frac{\alpha-1}{\alpha}f} + d_3 \odot \mathbf{1}_{03,07} + \ket{t^*_{03}}\bra{\mathbf{1}_{03}} + \ket{t^*_{47}}\bra{\mathbf{1}_{47}} + \ket{t_8^*}\bra{\mathbf{1}_8}, t\Big\rangle \\
	 + \langle \tau_A, \sigma_A \rangle  + \sum_i \gamma_i (\Psihat(\sigma_i) - \langle \nabla \Psihat(\sigma_i), \sigma_i\rangle) + \sum_x \lambda_x \langle x|\sigma_A|x\rangle \\
\end{multline}
where $\mathbf{1}_{x,y}$ is a matrix that is one on the ranges $x,y$ and zero otherwise. We can now eliminate them using standard conic duality, obtaining the dual problem

 \begin{equation}
 \begin{gathered}
 \min_{\omega^*, \tau_A,a_1, a_2, a_3, b_1, b_2, b_3, c_1, c_2, c_3, g^*, \gamma, \lambda, \mu, \nu, d_1, d_2, d_3, t_8^*, t_{03}^*, t_{47}^*}  \langle \tau_A, \sigma_A \rangle  + \sum_i \gamma_i (\Psihat(\sigma_i) - \langle \nabla \Psihat(\sigma_i), \sigma_i\rangle) + \sum_x \lambda_x \langle x|\sigma_A|x\rangle \\
 \text{s.t.}\quad\omega^* \succeq 0, \tau_A \succeq 0\\
 \left(a_1(x,z),a_2(x,z),a_3(x,z)\right) \in\mc K_\mathrm{rs}, \; x \in \{0,\ldots,3\} , \; z \in \{4,\ldots,7\} \\
(d_1(x,z), d_2(x,z), d_3(x,z) \in \mc{K}_{rs}, x \in \{0,\ldots,3\} , \; z \in \{0,\ldots,7\} \\
 \left(b_1, b_2, b_3 \right)\in \mc{K}^{*,1/\alpha}_{pc}, \left(c_1, c_2, c_3\right)\in \mc K^{*,1/\alpha}_{pc} \\
 g^* \ge 0, \gamma \ge 0, \mu \ge 0, \nu \ge 0 \\
\sum_{c \in \tilde{\mc{C}}} \bar{p}(c) 2^{\frac{\alpha-1}{\alpha}f (c)}\hat{\Pi}^c - \pkey 2^{\frac{\alpha-1}{\alpha} f ({\perp})} R^\top - \tau_A \otimes \id_B + \sum_{x =0}^3\sum_{z=4}^7a_1(x,z) \frac{1}{2} \hat\Pi^{(x,z)} + (b_2 + c_1 + c_2) R^\top \\ 
     + \sum_i \gamma_i \nabla\Psihat(\sigma_i) - \sum_x \lambda_x \proj{x} \otimes \id_B + \sum_{x=0}^3\sum_{z=0}^7d_1(x,z)\frac{1}{2} \hat\Pi^{(x,z)} + \omega^* = 0 \\
	  \sum_i \gamma_i -\pkey 2^{\frac{\alpha-1}{\alpha} f ({\perp})} = 0 \\
	 c_3 + \pkey 2^{\frac{\alpha-1}{\alpha} f ({\perp})} = 0 \\
	 b_1 - c_1 = 0 \\
	 g^* + a_3 + (b_2 + b_3 + c_1|\hat{\mc{Z}}|^{\alpha-1} + c_2) \mathbf{1} = 0 \\
\bar{p} \odot 2^{\frac{\alpha-1}{\alpha}f} + a_2\odot\mathbf{1}_{03,47} +d_2\odot\mathbf{1}_{03,07} + (b_2 + c_1 + c_2)\mathbf{1}_{03,47}- \ket{\lambda}\bra{\mathbf{1}_{08}}  \\ 
	 +\left(\sum_{z=0}^3\bra{N_c+1}\tilde R_B^z\ket{N_c+1}\right)\ket{\mu}\bra{\mathbf{1}_{47}} -\bra{N_c+1}R_B^\top\ket{N_c+1}\ket{\mu}\bra{\mathbf{1}_{03}} \\
+\bra{N_c+1}R_B^\top\ket{N_c+1}\ket{\nu}\bra{\mathbf{1}_8} - \bra{N_c+1}\tilde R_B^8\ket{N_c+1} \ket{\nu}\bra{\mathbf{1}_{47}}= 0 \\
2\bar{p} \odot 2^{\frac{\alpha-1}{\alpha}f} + d_3 \odot \mathbf{1}_{03,07} + \ket{t^*_{03}}\bra{\mathbf{1}_{03}} + \ket{t^*_{47}}\bra{\mathbf{1}_{47}} + \ket{t_8^*}\bra{\mathbf{1}_8} = 0
 \end{gathered}
 \end{equation}
which can be simplified to
 \begin{equation}
 \begin{gathered}
 \min_{\tau_A,a_1, a_2, a_3, b_1, b_2, b_3,  c_2, \gamma, \lambda, \mu, \nu, d_1, d_2, d_3, t_8^*, t_{03}^*, t_{47}^*} \langle \tau_A, \sigma_A \rangle  + \sum_i \gamma_i (\Psihat(\sigma_i) - \langle \nabla \Psihat(\sigma_i), \sigma_i\rangle) + \sum_x \lambda_x \langle x|\sigma_A|x\rangle \\
 \text{s.t.}\quad\tau_A \succeq 0\\
 \left(a_1(x,z),a_2(x,z),a_3(x,z)\right) \in\mc K_\mathrm{rs}, \; x \in \{0,\ldots,3\} , \; z \in \{4,\ldots,7\} \\
(d_1(x,z), d_2(x,z), d_3(x,z) \in \mc{K}_{rs}, x \in \{0,\ldots,3\} , \; z \in \{0,\ldots,7\} \\
 \left(b_1, b_2, b_3 \right)\in \mc{K}^{*,1/\alpha}_{pc}, \left(b_1, c_2, -\pkey 2^{\frac{\alpha-1}{\alpha} f ({\perp})}\right)\in \mc K^{*,1/\alpha}_{pc} \\
 \gamma \ge 0, \mu \ge 0, \nu \ge 0 \\
-\sum_{c \in \tilde{\mc{C}}} \bar{p}(c) 2^{\frac{\alpha-1}{\alpha}f (c)}\hat{\Pi}^c - \sum_{x =0}^3\sum_{z=4}^7a_1(x,z) \frac{1}{2} \hat\Pi^{(x,z)}   - \sum_{x=0}^3\sum_{z=0}^7d_1(x,z)\frac{1}{2} \hat\Pi^{(x,z)}  \\ 
    + (\pkey 2^{\frac{\alpha-1}{\alpha} f ({\perp})} - b_2 - b_1 - c_2) R^\top + (\tau_A + \operatorname{diagm}(\lambda)) \otimes \id_B  - \sum_i \gamma_i \nabla\Psihat(\sigma_i)\succeq 0 \\
	 \sum_i \gamma_i -\pkey 2^{\frac{\alpha-1}{\alpha} f ({\perp})} = 0 \\
	 -a_3 - (b_2 + b_3 + b_1|\hat{\mc{Z}}|^{\alpha-1} + c_2) \mathbf{1} \ge 0 \\
	\bar{p} \odot 2^{\frac{\alpha-1}{\alpha}f} + a_2\odot\mathbf{1}_{03,47} +d_2\odot\mathbf{1}_{03,07} + (b_2 + b_1 + c_2)\mathbf{1}_{03,47}- \ket{\lambda}\bra{\mathbf{1}_{08}} \\ 
	 +\left(\sum_{z=0}^3\bra{N_c+1}\tilde R_B^z\ket{N_c+1}\right)\ket{\mu}\bra{\mathbf{1}_{47}} -\bra{N_c+1}R_B^\top\ket{N_c+1}\ket{\mu}\bra{\mathbf{1}_{03}} \\
+\bra{N_c+1}R_B^\top\ket{N_c+1}\ket{\nu}\bra{\mathbf{1}_8} - \bra{N_c+1}\tilde R_B^8\ket{N_c+1} \ket{\nu}\bra{\mathbf{1}_{47}}= 0 \\
2\bar{p} \odot 2^{\frac{\alpha-1}{\alpha}f} + d_3 \odot \mathbf{1}_{03,07} + \ket{t^*_{03}}\bra{\mathbf{1}_{03}} + \ket{t^*_{47}}\bra{\mathbf{1}_{47}} + \ket{t_8^*}\bra{\mathbf{1}_8} = 0
 \end{gathered}
 \end{equation}
\end{document}